\documentclass[aps,prb,twocolumn,superscriptaddress,notitlepage]{revtex4-1}
\usepackage{graphicx}
\usepackage[caption=false]{subfig}
\usepackage{float}
\usepackage{natbib}
\usepackage{xcolor}
\usepackage{xr}
\usepackage{amsmath}
\usepackage{amssymb}
\usepackage{array}
\usepackage{wasysym}
\usepackage{color,soul}
\usepackage{braket}
\usepackage{verbatim}
\usepackage{hyperref}
\usepackage{gensymb}
\usepackage{url}
\usepackage{tabularx}
\usepackage{filecontents}
\usepackage{textgreek}
\begin{document}

\title{One-Dimensional Quantum Magnetism in the $S=1/2$ Mo(V) system, KMoOP\textsubscript{2}O\textsubscript{7}}

\author{Aly~H.~Abdeldaim}
\email{axa1699@student.bham.ac.uk}
\affiliation{School of Chemistry, University of Birmingham, Edgbaston, Birmingham B15 2TT, United Kingdom}
\affiliation{ISIS Neutron and Muon Source, Science and Technology Facilities Council, Didcot OX11 0QX, United Kingdom}
\author{Alexander~A.~Tsirlin}
\affiliation{Felix Bloch Institute for Solid-State Physics, Leipzig University, 04103 Leipzig, Germany}
\author{Jacques~Ollivier}
\affiliation{Institut Laue-Langevin, 71 Avenue des Martyrs, CS20156, 38042 Grenoble Cédex 9, France}
\author{Clemens~Ritter}
\affiliation{Institut Laue-Langevin, 71 Avenue des Martyrs, CS20156, 38042 Grenoble Cédex 9, France}
\author{Dominic~Fortes}
\affiliation{ISIS Neutron and Muon Source, Science and Technology Facilities Council, Didcot OX11 0QX, United Kingdom}
\author{Robin~S.~Perry}
\affiliation{London Centre for Nanotechnology and Department of Physics and Astronomy,University College London, London WC1E 6BT, United Kingdom}
\affiliation{ISIS Neutron and Muon Source, Science and Technology Facilities Council, Didcot OX11 0QX, United Kingdom}
\author{Lucy~Clark}
\affiliation{School of Chemistry, University of Birmingham, Edgbaston, Birmingham B15 2TT, United Kingdom}
\author{Gøran~J.~Nilsen}
\email{goran.nilsen@stfc.ac.uk}
\affiliation{ISIS Neutron and Muon Source, Science and Technology Facilities Council, Didcot OX11 0QX, United Kingdom}
\affiliation{Department of Mathematics and Physics, University of Stavanger, 4036 Stavanger, Norway}
\date{\today}

\begin{abstract}
\noindent
We present a comprehensive experimental and \textit{ab-initio} study of the $S=1/2$ Mo\textsuperscript{5+} system, KMoOP$_2$O$_7$, and show that it realizes the $S = 1/2$ Heisenberg chain antiferromagnet model. Powder neutron diffraction reveals that KMoOP$_2$O$_7$ forms a magnetic network comprised of pairs of Mo\textsuperscript{5+} chains within its monoclinic $P2_1/n$ structure. Antiferromagnetic interactions within the Mo\textsuperscript{5+} chains are identified through magnetometry measurements and confirmed by analysis of the magnetic specific heat. The latter reveals a broad feature centred on $T_\textrm{N} = 0.54$ K, which we ascribe to the onset of long-range antiferromagnetic order. No magnetic Bragg scattering is observed in powder neutron diffraction data collected at 0.05 K, however, which is consistent with a strongly suppressed ordered moment with an upper limit $\mu_\textrm{ord} < 0.15$ \textmu$_\textrm{B}$. The one-dimensional character of the magnetic correlations in KMoOP$_2$O$_7$ is verified through analysis of inelastic neutron scattering data, resulting in a model with $J_\textrm{1} \approx 34$ K and $J_\textrm{2} \approx -2$ K for the intrachain and interchain exchange interactions, respectively. The origin of these experimental findings are addressed through density-functional theory calculations.  
\end{abstract}
\maketitle
\section{Introduction}
\noindent The relevance of the dynamical properties of fractionalized states of matter to fault-tolerant quantum computing\cite{Nayak2008,Kitaev2003,Lahtinen2017} demands the realization of such quantum states in strongly correlated electron systems\cite{Broholm2020}. This interest is complemented by the need to develop a fundamental understanding of fractionalized states in condensed matter systems, such as certain quantum spin liquid models, that host fractionalized quasiparticle excitations\cite{Broholm2020,Savary2016,Clark2021}. The broad interest in the field is highlighted by recent theoretical and experimental investigations of the emergent fractional quasiparticles of spin models hosting, to name a few, monopoles\cite{Castelnovo2012}, Majorana fermions\cite{Hermanns2018}, and the fractional quantum Hall effect\cite{Feldman2021}. 

In this vein, owing to the exact solvability of its spin Hamiltonian\cite{Bethe1931, Hulthen1938}, the $S = 1/2$ Heisenberg chain antiferromagnet model (HAF) offers a canonical platform for the direct observation of fractional excitations. Known as spinons, the elementary excitations of the HAF manifest as continua as observed in the landmark neutron scattering studies of KCuF$_3$\cite{Tennant1993} and CuSO$_4\cdot5$H$_2$O\cite{Mourigal2013}. More recently, focus has shifted towards investigating the ground states arising from the perturbative effects of spin-orbit coupling, applied magnetic fields, and frustration to the HAF model\cite{Coldea, Wang2015,Vasiliev2018}. A less explored model in this context is the frustrated $S = 1/2$ Heisenberg chain model (FCM), which provides a rich magnetic phase diagram dependent on the degree of frustration, $\alpha=J_\textrm{2}/J_\textrm{1}$, between the intrachain, $J_\textrm{1}$, and interchain, $J_\textrm{2}$, exchange interactions. Although a spontaneous dimer-fluid ground state has been predicted for $\alpha > 0.241$\cite{Haldane1982,Okamoto1992,Eggert1996}, only a few experimental realizations have been identified for the FCM model\cite{Enderle2005,Dutton2012,Nilsen2008,Kasinathan2013}, and the presence of the spin gap associated with the dimerization has yet to be conclusively observed.

\begin{figure*}
\centering
\includegraphics[width = \textwidth,keepaspectratio]{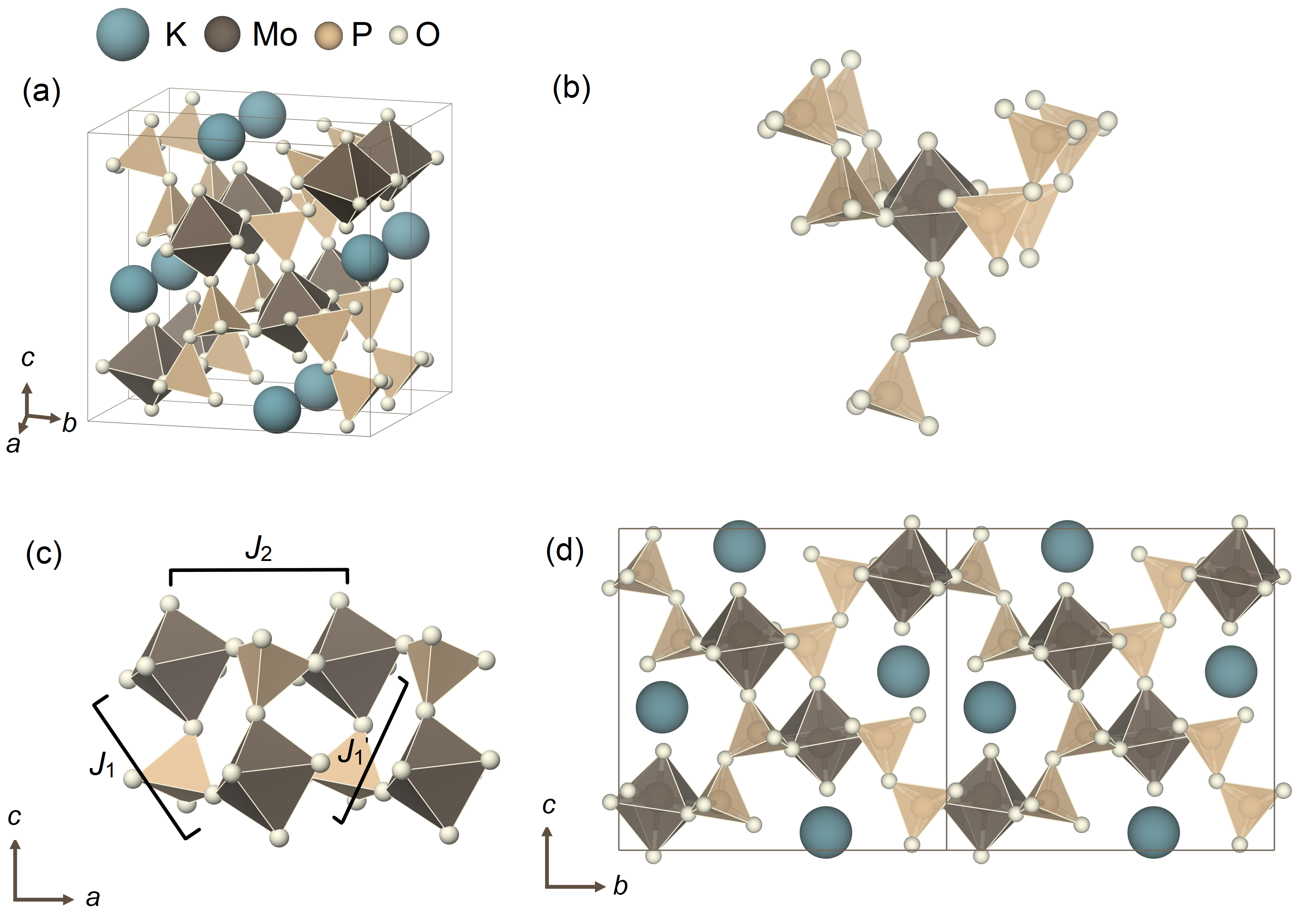}\caption{The $P2_1/n$ crystal structure of KMoOP$_2$O$_7$ as viewed along the [111] direction (b) the Mo$^{5+}$ ions are situated within a distorted octahedral environment and are bridged by [PO$_4$]$^{3-}$ tetrahedra along the crystallographic \textit{a}-axis to form (c) a quasi-one-dimensional network of pairs of Mo(V) chains with possible intrachain, $J_2$, and interchain, $J_1$ and $J_1'$, superexchange Mo-O-P-O-Mo interaction pathways. (d) This forms a tunnel structure along the crystallographic \textit{a}-axis where the K$^{+}$ ions lie. Figure generated using the \texttt{VESTA} visualization software\cite{Momma:ko5060}}
\label{Structure}
\end{figure*}

Motivated by the richness of the exotic phases predicted for this model, we have explored the Mo$^{5+}$ pyrophosphate family of materials, $A$MoOP$_2$O$_7$ ($A$ = Na-Cs)\cite{Gueho1992,Canadell1997,Ledain1996,Guesdon1994}, whose structures form pairs of chains of 4$d^1$ Mo\textsuperscript{5+} ions (Fig.\ref{Structure}) that could be amenable to frustration. To this end, we here investigate the structural and magnetic properties of KMoOP$_2$O$_7$ (Fig.\ref{Structure}) and show that it falls on the Heisenberg one-dimensional limit of the $S = 1/2$ HAF model. We begin by reporting our experimental and \textit{ab-initio} methodology in Sec.\ref{sec:methods}. Using neutron powder diffraction, we verify the previously published monoclinic structure\cite{Gueho1992} at room temperature and present its temperature dependence down to 50 mK (Sec.\ref{sec:res-cryst}). The one-dimensional magnetic behavior of the system is then explored through thermodynamic property measurements and neutron powder diffraction (NPD) (Sec.\ref{sec:res-mag}-\ref{sec:res-LRO}). Following this, the one-dimensional character of the magnetic correlations in KMoOP$_2$O$_7$ are discussed through the lens of its electronic structure in Sec.\ref{sec:mmm}. We finally estimate the parameters of the Hamiltonian using inelastic neutron scattering (INS) in Sec.\ref{sec:res-INS} before concluding in Sec.\ref{sec:conc}.
\vspace{-2mm}
\section{Methods}
\label{sec:methods}
%%%%%%%%%%%%%%%%%%%%%%%%%%%%%%%%%%%%%
%%%%%%%%%%%%%%%%%%%%%%%%%%%%%%%%%%%%%
\noindent Polycrystalline samples of KMoOP$_2$O$_7$ were prepared following a modified version of a previously published method\cite{Gueho1992}. MoO$_3$ (Alfa Aesar, 99.998\%), Mo (powder, $<$150 μm, 99.9\%) (NH$_4$)$_2$HPO$_4$ (Alfa Aesar, 98\%+), and K$_2$CO$_3$ (Sigma Aldrich, 99.995\%) were combined in the molar ratio 1.72:0.28:4:1 and intimately ground in a planetary ball mill for 30 minutes in an isopropanol medium and then pressed into a pellet. Samples were sintered in an alumina crucible at 773~K for 24 hours before being reground and pelletized for a second heating stage at 973~K for 24 hours that resulted in a green colored product. A 5 g sample of this product was used in all of the following experiments. The synthesis of the isostructural diamagnetic analogue, KNbOP$_2$O$_7$, is described in the supplementary material.

Powder neutron diffraction experiments were carried out in 10 K steps between 300~K and 200~K on the high-resolution powder diffractometer (HRPD) at the ISIS Neutron and Muon Source. To probe the possible presence of magnetic scattering, powder neutron diffraction experiments were also performed on the D20 diffractometer\cite{D20} at the Institut Laue-Langevin (ILL) and on the WISH diffractometer at the ISIS Neutron and Muon Source. Two data sets were collected on each instrument at 1.8 K and 0.05 K and 20 K and 1.8 K, respectively. The structural model was refined using the Rietveld method applied in the \texttt{GSAS} software package\cite{Toby2001}. Magnetic structure analysis was conducted using the \texttt{Mag2Pol} software package\cite{Qureshi2019}. Similar to the approach taken for other Mo\textsuperscript{5+}-containing materials\cite{Ishikawa2017}, we use an averaged Cr\textsuperscript{4+} and W\textsuperscript{5+} magnetic form factor to approximate that of Mo\textsuperscript{5+}.

Temperature dependent DC magnetic susceptibility data were measured on a Quantum Design MPMS3 SQUID magnetometer between 1.8 K and 300 K using a 22.25 mg sample. Measurements were performed using both zero-field- and field-cooled protocols in an applied magnetic field of 1000~Oe. As no splitting between the zero-field- and field-cooled curves was observed, only the zero-field-cooled measurement is shown here. Calculations for the exact diagonalization of the Heisenberg frustrated chain model were performed using the \texttt{ALPS} software package on a chain of $L = 18$ spins. The resulting model was fit to the experimental data using non-linear least squares regression, where the only fitting parameter was the intrachain exchange interaction, $J_2$. The interchain coupling, $J_1$, was varied in 0.01$J_2$ steps between $-0.3J_2$ and $0.3J_2$. 

Specific heat measurements were performed on a 5.58~mg sample on a Quantum Design PPMS measurement system between 1.8~K and 300~K. A lower temperature measurement, between 0.1~K and 4~K, was performed using a dilution refrigeration insert. 

The dynamical structure factor, $S\text{(}Q,E=\hbar \omega \text{)}$, was measured on the direct geometry time-of-flight spectrometer IN5 at the ILL\cite{IN5}. An incident energy of $E_\textrm{i} = 14.2$ meV (2.4 \AA) was used to collect data at 1.8 K for both KMoOP$_2$O$_7$ and its isostructural diamagnetic analogue, KNbOP$_2$O$_7$. The exact expression for the two- and four-spinon continuum, $S\textrm{(}Q,E\textrm{)}^{\textrm{calc}}_\textrm{2+4}$, was obtained from the exact expression of Caux and Hagemans\cite{Caux2006}. To determine the powder-averaged $S\textrm{(}Q,E\textrm{)}^{\textrm{calc}}_\textrm{2+4}$, a uniform cross-section of the constant-$Q$ sphere was obtained by normalizing random coordinates generated by a Gaussian distribution to the sphere radius. The effect of the interchain exchange interaction, $J_\textrm{2}$, on the other hand, was calculated using the random phase approximation (RPA)-style approach developed by Kohno et al.\cite{Kohno2007}. Fitting the experimental dynamical structure factor using this model was done using a particle swarm optimization algorithm. Here, the fitting parameters were constant across all four cuts used and were the intra- and inter-chain exchange parameters $J_\textrm{2}$ and $J_\textrm{1}$, an amplitude, and a background term. 

Interaction parameters of the spin Hamiltonian
\begin{equation}
    \mathcal{H} = \sum_{\langle ij \rangle}J_{ij}S_iS_j
\end{equation}
where the summation is over lattice bonds $\langle$ij$\rangle$, were obtained from density-functional theory (DFT) band-structure calculations performed in the \texttt{FPLO} \cite{Koepernik1999} code using generalized gradient approximation (GGA) for the exchange-correlation potential \cite{Perdew1996}. A dense $k$-mesh with up to 152 points in the symmetry-irreducible part of the Brillouin zone for the crystallographic unitcell and 64 points in the doubled supercell were used. All calculations were performed in the full-relativistic mode for the experimental crystal structure of KMoOP$_2$O$_7$ determined in this work.

The exchange parameters $J_{\textrm{ij}}$ were extracted using two complementary approaches. In the first, we calculated hopping integrals between the Mo 4$d$ states using Wannier functions constructed for the uncorrelated (GGA) band structure, and introduced these hoppings into the Kugel-Khomskii model that leads to the magnetic exchange couplings as follows \cite{Mazurenko2006,Tsirlin2011},
\begin{equation}
J_\textrm{ij} = \frac{4[t_\textrm{ij}^{(nn)}]^2}{U_\textrm{eff}}-\sum_{m}\frac{4[t_\textrm{ij}^{(nm)}]^2J_\textrm{eff}}{(U_\textrm{eff}+\Delta_\textrm{m})(U_\textrm{eff}-J_\textrm{eff}+\Delta_\textrm{m})},  
\label{eqn2}
\end{equation}
where the first and second terms represent antiferromagnetic and ferromagnetic contributions, respectively. Here, $U_\textrm{eff} = 4$ eV is the effective on-site Coulomb repulsion and $J_\textrm{eff} = 0.5$ eV is the effective Hund's coupling in the Mo 4$d$ shell. The hoppings $t_\textrm{ij}^{(\textrm{nn})}$ are between the half-filled states of Mo, whereas $t_\textrm{ij}^{(\textrm{nm})}$ involve the higher-lying empty states $m$, and $\Delta_\textrm{m} = \epsilon_\textrm{m} - \epsilon_\textrm{n}$ is the crystal-field splitting.
In the second approach, we obtained the exchange couplings by a mapping procedure \cite{Xiang2011, Tsirlin2014} using total energies of collinear spin configurations evaluated within DFT+\textit{U}, where correlation effects in the Mo 4$d$ shell are treated on a mean-field level with the on-site Coulomb repulsion $U_\textrm{d} = 4$ eV, Hund's coupling, $J_\textrm{d} = 0.5$ eV, and atomic-limit flavor of the double-counting correction. This set of parameters gave the best agreement with the experimental results for the leading exchange coupling, although the corresponding $\tilde U = U_\textrm{d}-J_\textrm{d} = 3.5$ eV is somewhat higher than the 2.0 - 2.5 eV used in previous studies \cite{Iqbal2017,Hembacher2018,Abdeldaim2019}. The DFT+\textit{U} calculations were performed for the supercell doubled along the \textit{a}-axis due to strong antiferromagnetic interactions along this direction.

%%%%%%%%%%%%%%%%%%%%%%%%%%%%%%%%%%
\begin{figure}
\centering
\includegraphics[width=0.48\textwidth,height=0.48\textheight,keepaspectratio]{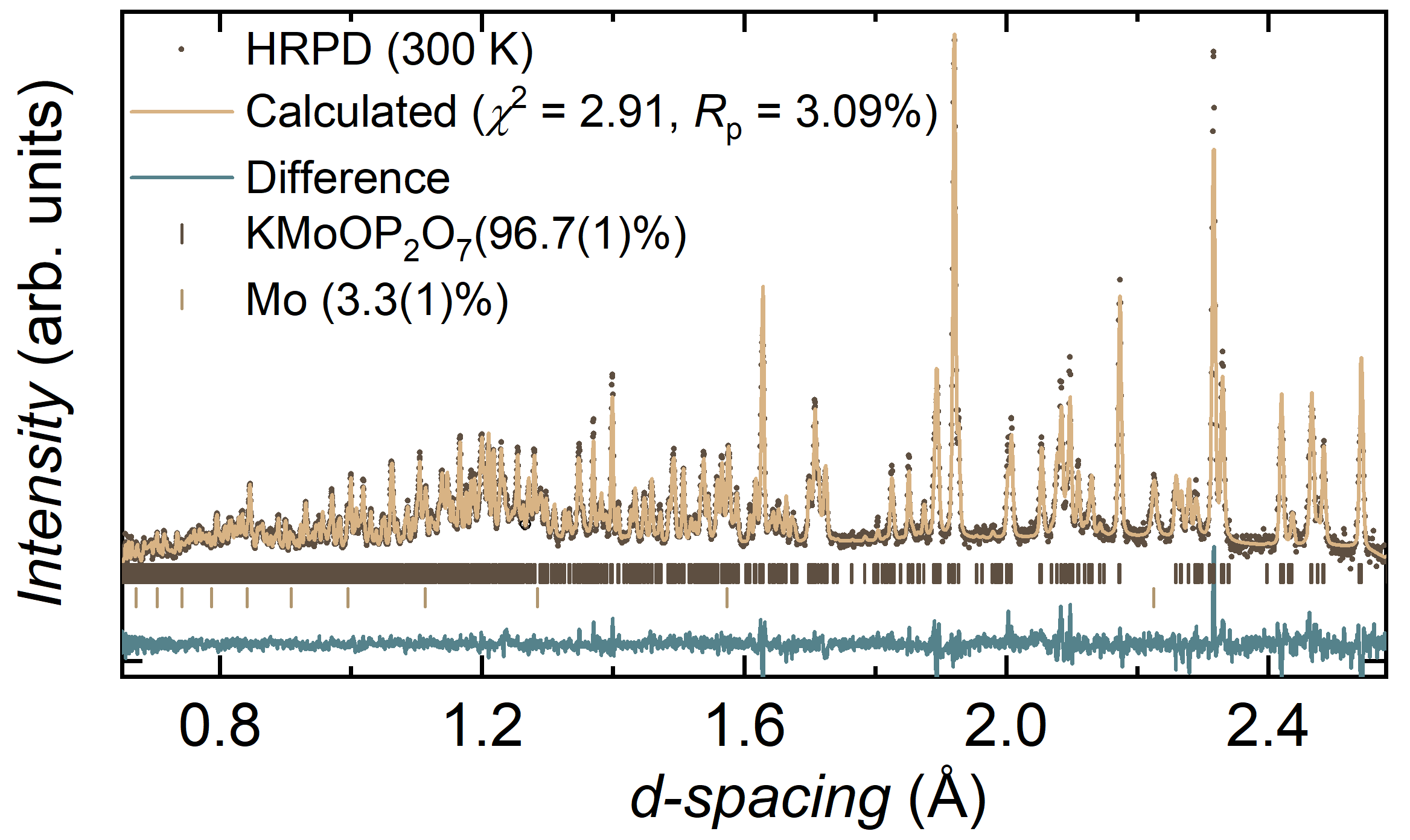}\caption{Rietveld refinement of the $P2_1/n$ model to neutron powder diffraction data collected at 300 K on the HRPD instrument at the ISIS Neutron and Muon Source with goodness-of-fit parameters $\chi^2 = 2.91$ and $R_\textrm{p} = 3.09$\% and the structural model $a = 5.0842(1)$ \AA, $b = 11.7243(2)$ \AA, $c = 11.5077(2)$ \AA, $\beta = 90.975(1) \degree$.}
\label{HRPD}
\end{figure}
%%%%%%%%%%%%%%%%%%%%%%%%%%%%%%%%%%%%%
\section{Results}
\subsection{Crystal structure}
\label{sec:res-cryst}
\noindent
Through Rietveld analysis of NPD data, the $P2_1/n$ model \cite{Gueho1992} was confirmed to describe the crystal structure of KMoOP$_2$O$_7$ at all measured temperatures down to 50 mK. A representative Rietveld refinement ($\chi^2 = 2.91$, $R_\textrm{p} = 3.09\%$) of data collected at 300 K, which results in the model described in Table S.1, is shown in Fig.\ref{HRPD}. Analysis of data collected at different temperatures is shown in Fig. S2. The resulting crystal structure (Fig.\ref{Structure}(a)) can be described as a network of octahedrally coordinated Mo$^{5+}$ ions (Fig.\ref{Structure}(b)) that propagate along the crystallographic $a$-axis through [PO$_4$]$^{3-}$ tetrahedral bridges to form Mo$^{5+}$-containing chains. Each Mo-octahedron is coordinated by five P$_2$O$_7$ groups which leave a short apical bond, connected to a K cation, that distorts the octahedral geometry (Fig.\ref{Structure}(c)). This forms a tunnel like cavity, occupied by K cations, that lie along the crystallographic $a$ axis (Fig.\ref{Structure}(d)). The relevant nearest-neighbour superexchange interactions are most likely to be mediated by the Mo-O-P-O-Mo pathways within and between the pairs of chains. Following the nomenclature of the Heisenberg frustrated chain model, those interactions can be described by the intrachain, $J_\textrm{2}$, and interchain, $J_\textrm{1}$ and $J_\textrm{1}^{'}$, superexchange parameters with Mo-Mo distances of 5.09(1) \AA, 5.36(1) \AA, and 5.40(1) \AA, respectively (Fig.\ref{Structure}(c)). These pairs of chains are separated from others within the bc plane by pyrophosphate molecules with distances ranging 6.18(1)-6.94(1) \AA. Similar to other Mo$^{5+}$ and V$^{4+}$ containing materials\cite{Canadell1997.Mukharjee2021,Urushihara2020}, the formation of a short apical bond distorts the octahedral geometry with apical bond distances of $d_\textrm{Mo-O7} = 1.67(1)$ \AA~and $d_\textrm{Mo-O1} = 2.14(1)$ \AA.

\subsection{Magnetometry}
\label{sec:res-mag}
\begin{figure}
\centering
\includegraphics[width=0.45\textwidth,height=0.45\textheight,keepaspectratio]{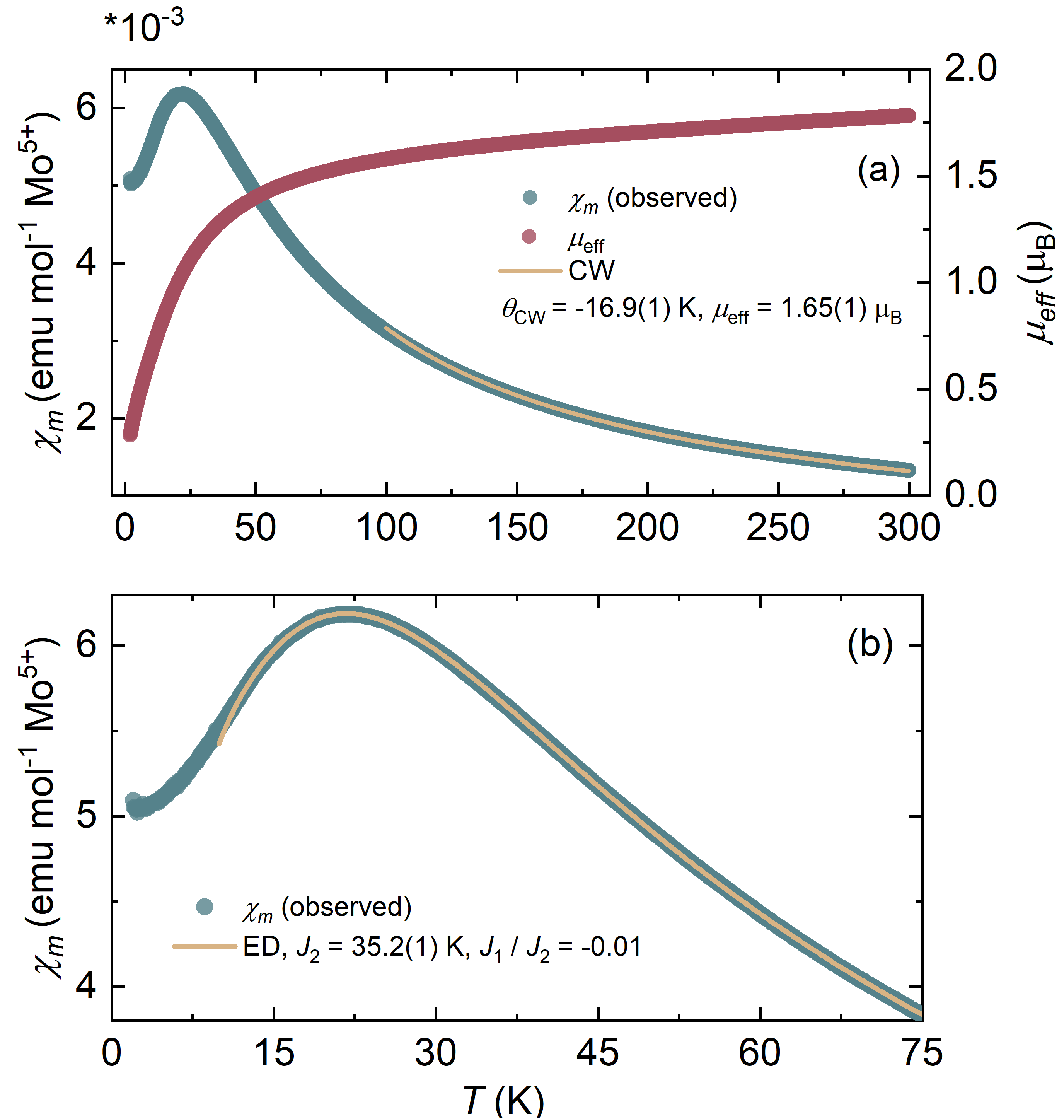}\caption{(a,left) Temperature dependent magnetic susceptibility measured in an applied field of $H = 1000$ Oe alongside the (a, right) calculated effective moment, $\mu_{\textrm{eff}}$. A Curie-Weiss fit to the data yields an antiferromagnetic Weiss constant $\theta_{\textrm{CW}} = -16.9(1)$ K and an effective moment, $\mu_{\textrm{eff}} = 1.65(1)$ $\mu_{\textrm{B}}$, reflecting a nearly full spin-only $S = 1/2$ moment. (b) Fitting the magnetic susceptibility to an exact diagonalization (ED) calculation of the Heisenberg frustrated chain model with an interchain exchange parameter, $J_1$, and an intrachain exchange, $J_2$, yields an effective one-dimensional model with $J_2 = 35.2(1)$ K and $J_1 / J_2 = -0.01$.}

\label{magnetometry}

\end{figure}

\noindent  
The temperature dependence of the zero-field cooled molar magnetic susceptibility is shown in Fig.\ref{magnetometry} (a). Between 160 and 300 K, $\chi_\textrm{m}$ is well described by the modified Curie-Weiss (CW) law, $\chi_\textrm{m} (T) = C/(T-\theta_\textrm{CW}) + \chi_\textrm{0}$, where $C = N_\textrm{A}\mu_\textrm{eff}^2/3k_\textrm{B}$  and $\theta_\textrm{CW}$ are the Curie and Weiss constants, respectively, and $\chi_\textrm{0}$ is a temperature independent background term. Here, the choice for the minimum fitting temperature of 170 K was made based on the stabilization of the extracted parameters beyond this temperature (Fig. S3). The fit yields the parameters $\theta_\textrm{CW} = -16.9(1)$~K, $C = 0.341(1)$~emu~K~mol\textsuperscript{-1} ($g = 1.91)$ , $\mu_\textrm{eff} = 1.65(1)$ \textmu$_\textrm{B}$, and $\chi_\textrm{0} = 2.51 \times~10^{-5}$~emu~mol\textsuperscript{-1}, suggesting dominant antiferromagnetic interactions and a moment size close to the full spin-only $S = 1/2$ moment of 1.73 \textmu$_\textrm{B}$. This is unlike other Mo$^{5+}$-containing materials such as Ba$_2$YMoO$_6$\cite{deVries} and Lu$_2$Mo$_2$O$_5$N$_2$\cite{Clark2014} where a significant orbital contribution is evident by a large deviation of $\mu_\textrm{eff}$ from its spin-only value. On cooling, the development of short-range correlations is evidenced by a broad symmetrical feature, characteristic of quasi-one-dimensional materials, that is centered about 21~K. No features indicative of long-range magnetic order are observed down to 1.8~K. 

To approximate the leading magnetic exchanges in KMoOP$_2$O$_7$, an exact diagonalization calculation has been performed for the $S=1/2$ Heisenberg frustrated chain model using the \texttt{ALPS} software package. Given the similarity between the bonding geometries represented by $J_\textrm{1}$ and $J_\textrm{1\textsuperscript{'}}$, and thus the similarity between the likely superexchange mechanisms, only one interchain exchange is used to approximate both parameters. When fitting $\chi_m$ above 15 K, the resulting parameters ($J_\textrm{2} = 35.2(1)$ K, $J_\textrm{1} / J_\textrm{2} = -0.01$, $g$ = 1.96) are consistent with a leading antiferromagnetic interaction and highlight the one-dimensionality of the magnetism in KMoOP$_2$O$_7$. The minimum fitting temperature here was picked within a region where minimal variation in $\chi^2$ and $J_2$ is observed. In comparison to the CW model, $J_\textrm{1}$ could then be estimated using $\theta_\textrm{CW} = -\frac{1}{4} (zJ_\textrm{1}+z^{'}J_\textrm{2})$, where $z$ is the number of couplings per site, resulting in $J_\textrm{1}/J_\textrm{2} \approx -0.04$.  While the small $J_\textrm{1} / J_\textrm{2} = -0.01$ reflects that fitting the magnetic susceptibility cannot produce a reliable estimate of $J_\textrm{1}$, the resulting model, where interchain exchange plays a minimal role, is confirmed by our electronic structural calculations (Sec.\ref{sec:mmm}) which places KMoOP$_2$O$_7$ within the one-dimensional limit of the frustrated HAF phase diagram. As a one-dimensional spin chain, this reflects an expected saturation field, $H_\textrm{sat} = 2J_\textrm{1D}/g\mu_\textrm{B}$, of $\approx 53$~T. 

%%%%%%%%%%%%%%%%%%%%%%%%%%%%%%%%%%
%%%%%%%%%%%%%%%%%%%%%%%%%%%%%%%%%%%

\subsection{Specific Heat}
\label{sec:res-spec}
\begin{figure}
\centering
\includegraphics[width=0.48\textwidth,height=0.48\textheight,keepaspectratio]{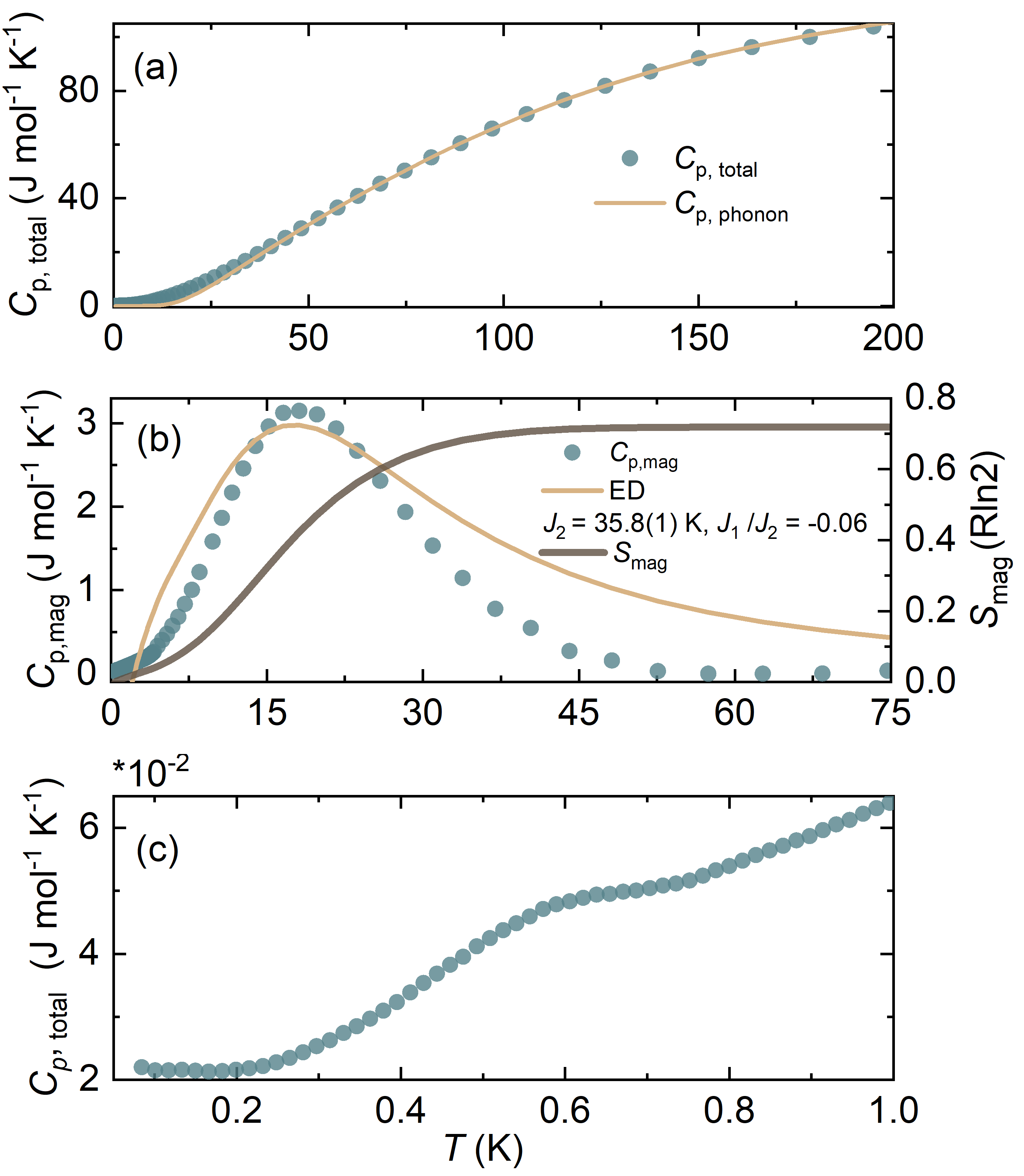}\caption{(a) Temperature dependence of the zero-field total heat capacity fit to a Debye-Einstein model to approximate the phonon contribution to the specific heat, $C_\textrm{p, phonon}$, yielding $\theta_\textrm{D} \simeq 550$ K, $f_\textrm{D} \simeq 0.58$, $\theta_\textrm{E\textsubscript{1}} \simeq 90$ K, $f_\textrm{E\textsubscript{1}} = 0.11$, $\theta_\textrm{E\textsubscript{2}} \simeq 115$ K, $f_\textrm{E\textsubscript{2}} = 0.13$, $\theta_\textrm{E\textsubscript{3}} \simeq 460$ K, and $f_\textrm{E\textsubscript{3}} = 0.18$ . By subtracting $C_\textrm{p, phonon}$ from the total heat capacity (b, left), the resulting magnetic specific heat, $C_\textrm{p, mag}$, reveals a broad feature centered about 18 K and that can be described by the S=1/2 Heisenebrg frustrated chain model with an intrachain exchange parameter, $J_\textrm{1} = 35.8(1)$ K, and an interchain exchange parameter, $J_\textrm{2} / J_\textrm{1} = -0.06$. This feature is also associated with (b, right) magnetic entropy release. (c) A broad feature, centered about $T_\textrm{N} = 0.54$ K is observed in $C_\textrm{p, total}$.}

\label{specificheat}

\end{figure}

\noindent
The temperature dependence of the zero-field molar specific heat, $C_\textrm{p, total}$, is shown in Fig.\ref{specificheat}. Considering the insulating behavior expected, the total specific heat, $C_\textrm{p, total}$, can be approximated by the individual contributions of its lattice, $C_\textrm{p, phonon}$, and magnetic, $C_\textrm{p, mag}$, components such that $C_\textrm{p, total} = C_\textrm{p, phonon} + C_\textrm{p, mag}$. To estimate $C_\textrm{p, phonon}$, the phenomenological Debye-Einstein model,

\begin{equation}
    C_\textrm{p, phonon}(T) = f_\textrm{D}C_\textrm{D}(\theta_\textrm{D}, T) + \sum_{i=1}^3 f_\textrm{{E\textsubscript{i}}} C_\textrm{{E\textsubscript{i}}}(\theta_\textrm{{E\textsubscript{i}}}, T)
\end{equation}

\begin{equation}
    C_\textrm{D}(\theta_\textrm{D}, T) = 9nR(\frac{T}{\theta_\textrm{D}})^3 \int_{0}^{\frac{\theta_\textrm{D}}{T}} \frac{x^4 e^x}{(e^x -1)^2} dx
\label{debye}
\end{equation}

\begin{equation}
     C_{E}(\theta_{E}, T) = 3nR(\frac{T}{\theta_\textrm{E}})^2\frac{e^{\theta_\textrm{E}/T}}{(e^{\theta_\textrm{E}/T}-1)^2}
\label{einstein}
\end{equation}
is used. Here, Eqns.\ref{debye},\ref{einstein} represent the Debye and Einstein terms and their characteristic temperatures, $\theta_\textrm{D}$ and $\theta_\textrm{E}$, respectively, whereas $f_\textrm{D}$ and $f_\textrm{E}$ are weighting factors, $R$ is the universal gas constant, and $n$ defines the number of atoms per formula unit. When fitting down to 40 K (Fig. \ref{specificheat}(a)), $C_\textrm{p, total}$ is well described by the parameters $\theta_\textrm{D} \simeq 550$ K, $f_\textrm{D} \simeq 0.58$, $\theta_\textrm{E\textsubscript{1}} \simeq 90$ K, $f_\textrm{E\textsubscript{1}} = 0.11$, $\theta_\textrm{E\textsubscript{2}} \simeq 115$ K, $f_\textrm{E\textsubscript{2}} = 0.13$, $\theta_\textrm{E\textsubscript{3}} \simeq 460$ K, and $f_\textrm{E\textsubscript{3}} = 0.18$. Here, four Einstein terms were initially used to reflect the three optical phonons observed in the measured dynamical structure factor (Fig.S6) plus any additional higher energy phonon branches. Fitting the lowest energy phonon, however, proved the fit unstable, and so only three terms were used. By subtracting an extrapolation of the estimated $C_\textrm{p, phonon}$ from $C_\textrm{p, total}$ down to 0.7 K, a broad feature, consistent with an anomaly seen in the Fisher specific heat (Fig.S3), $d\chi T/dT$, is observed in $C_\textrm{p, mag}$ (Fig.\ref{specificheat}(b)). When using the extracted $J_\textrm{1} \simeq 35$~K from the exact diagonalization model describing $\chi_\textrm{m} (T)$, we find that the temperature at which this maximum is observed, $T_\textrm{max} \simeq 18$~K, is consistent with the expected $0.48J_\textrm{1}\simeq 18$~K for the $S = 1/2$ one-dimensional HAF model\cite{Johnston2000}. The maximum magnetic specific heat for this feature, $C_\textrm{p, mag}^\textrm{max} = 3.12$ J mol\textsuperscript{-1} K\textsuperscript{-1}, is also in agreement with the expected $C_\textrm{p, mag}^\textrm{max} = 0.3497\times R \simeq 2.9$ J mol\textsuperscript{-1} K\textsuperscript{-1} for a uniform $S = 1/2$ Heisenberg chain antiferromagnet\cite{Bernu2001}.

To confirm the validity of this model, we then fit $C_\textrm{p, mag} (T)$ to the theoretical curves obtained from the exact diagonalization calculation of the $S = 1/2$ Heisenberg frustrated chain model and obtain $J_\textrm{1} = 35.8(1)$ K and $J_\textrm{2} / J_\textrm{1} = -0.06$ (Fig.\ref{specificheat}(b)). While the resulting model describes the peak position in $C_\textrm{p, mag} (T)$ and is consistent with minimal frustration, an overestimation of the phonon contribution is evident when compared to the theoretical curve. Indeed, this discrepancy is also observed when estimating the spin entropy release associated with this feature which was calculated as $S_\textrm{mag} (T)= \int_{1~\text{K}}^{100~\text{K}} C_\textrm{p, mag}/T$. The resulting $S_\textrm{mag} = 4.15$ J mol\textsuperscript{-1} K\textsuperscript{-1} reflects that only $\approx 70\%$ of the maximum allowed $Rln2$ entropy is released across this feature, which is lower than the theoretical entropy release for the $S = 1/2$ HAF model\cite{Johnston2000}.

Below 1 K, the temperature dependence of the total zero-field specific heat reveals a feature centered about $T_\textrm{N}$ = 0.54 K. Assuming that it represents the onset of long-range magnetic order, the absence of any significant interchain exchange interactions, $J_\perp < 0.19$~K, is suggested by applying the analytical expression $\lvert J_\perp \rvert \simeq  T_\textrm{N}/1.28\sqrt{\ln(5.8J_\textrm{1}/T_\textrm{N})}$\cite{Schulz1996} with $T_\textrm{N} < 0.6$ K. An upper limit on the ordered magnetic moment, $\mu_\textrm{ord}$, can then be approximated as $\mu_\textrm{ord} < 0.08$~\textmu$_\textrm{B}$ by using the expression $\mu_\textrm{ord} \simeq 1.017\sqrt{\lvert J_\perp \rvert /J_\textrm{1}}$\cite{Schulz1996}.

%%%%%%%%%%%%%%%%%%%%%%%%%%%%%%%%%%
%%%%%%%%%%%%%%%%%%%%%%%%%%%%%%%%%%%

\subsection{Long Range Magnetic Order}
\label{sec:res-LRO}
\noindent
\noindent
To investigate the nature of the long-range magnetic order, NPD data were collected at 0.05, 1.8, and 20 K on the D20 and WISH diffractometers. While at first glance features consistent with a propagation vector of $\kappa = $ 0 can be gleaned from the subtracted data sets, the simulated models for this propagation vector confirm that these features are instead artefacts arising from the subtraction (Fig. S4(a)). Thus, no features consistent with magnetic Bragg scattering are observed in the subtracted data sets. Instead, if we assume antiferromagnetic ordering along the chain direction, $\kappa = $ (1/2 0 0), as justified by the crystal structure (Sec.\ref{sec:res-cryst}) and the electronic structure (Sec.\ref{sec:mmm}), an upper limit of $\mu_\textrm{ord} < 0.15$ \textmu$_\textrm{B}$ can be estimated for all moment directions at 0.05 K (Fig. S4(b)). This strong suppression of the ordered moment is consistent with KMoOP$_2$O$_7$ residing close to the pure one-dimensional chain limit. Indeed, such a reduction is characteristic of the $S = 1/2$ antiferromagnetic Heisenberg chain spin model and is generally observed across its various material realizations\cite{Nilsen2015,Belik2005}.

\subsection{Microscopic Magnetic Model}
\label{sec:mmm}
\noindent
\noindent To better understand the Heisenberg one-dimensional character of the magnetic correlations observed thus far, we now turn to the electronic structure. In the absence of correlations, the GGA band structure of KMoOP$_2$O$_7$ is metallic. The bands between $-$0.5 and 4~eV have predominantly Mo 4$d$ origin (Fig.S5). The apparent Heisenberg nature of the spins can be understood as the structure of KMoOP$_2$O$_7$ features strongly distorted MoO$_6$ octahedra. Here, the short apical Mo-O bond splits the $t_\textrm{2g}$ states into the lower-lying $d_\textrm{xy}$ orbital and higher-lying, nearly degenerate $d_\textrm{yz}$ and $d_\textrm{xz}$ orbitals. This raises the orbital degeneracy of 4$d^1$ Mo$^{5+}$ and leads to purely Heisenberg spins. This is similar to the crystal-field splitting of V$^{4+}$ that also forms short apical bonds within the VO$_6$ octahedra as in $\alpha$-KVOPO$_4$\cite{Mukharjee2021}. Orbital energies for KMoOP$_2$O$_7$ determined from the tight-binding fit of the band structure are $\epsilon_\textrm{xy} = 0.02$ eV,
$\epsilon_\textrm{yz} = 0.98$ eV, $\epsilon_\textrm{xz} = 1.03$ eV, $\epsilon_\textrm{x\textsuperscript{2}-y\textsuperscript{2}} = 3.09$ eV, and $\epsilon_\textrm{3z\textsuperscript{2}-r\textsuperscript{2}} = 4.50$ eV. The crystal-field splittings are about twice larger than in the case of V$^{4+}$ in an oxide crystal field\cite{Tsirlin2011,Tsirlin2011_2}, reflecting the larger spatial extent of 4$d$ orbitals compared to 3$d$.

%%%%%%%%%%%%%%%%%%%%%%%%%%%%%%%%%%%%%
\begin{figure}
\centering
\includegraphics[width = 0.65\columnwidth,keepaspectratio]{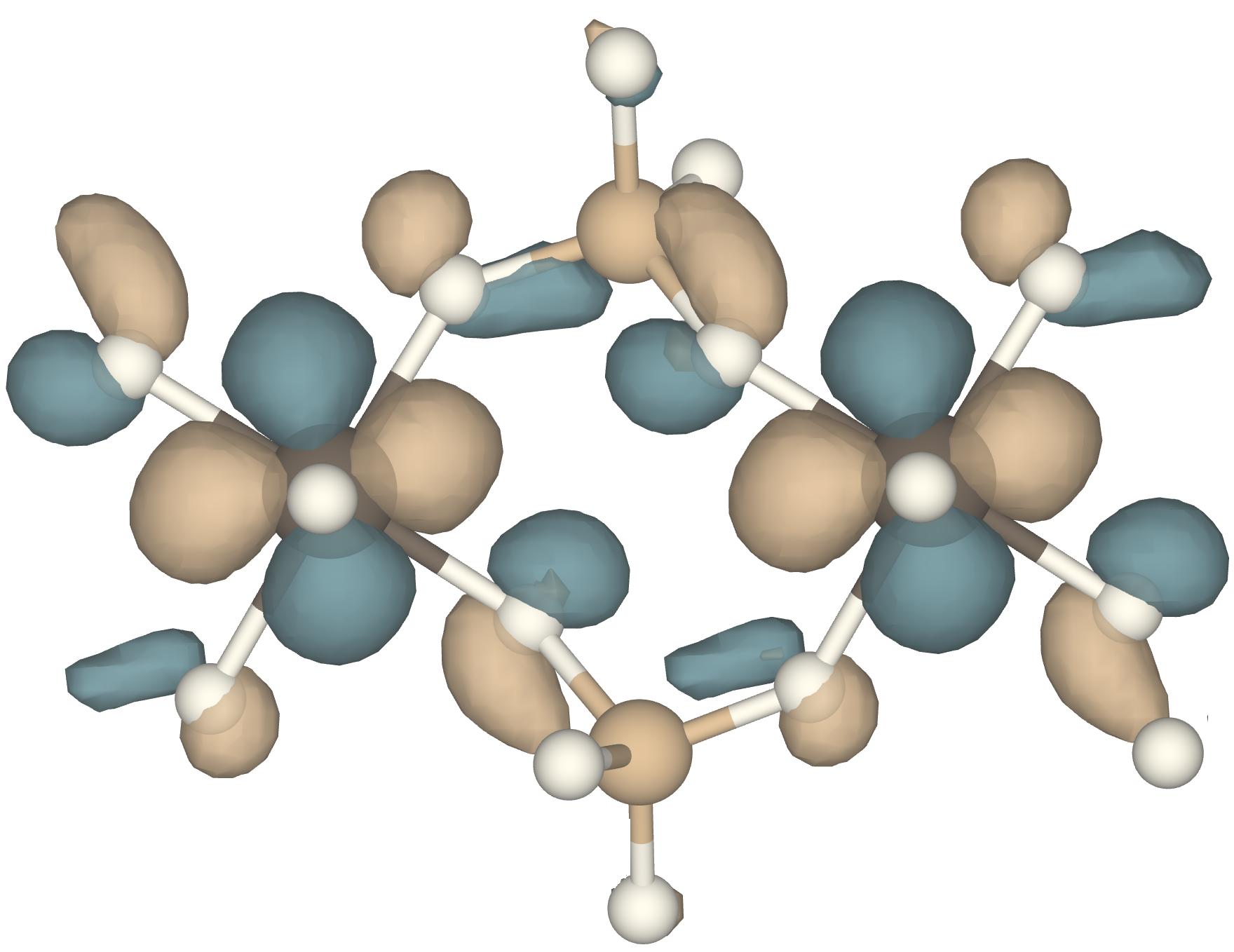}\caption{Mo $d_\textrm{xy}$-based Wannier functions showing the orbital overlap for the $J_\textrm{1}$ superexchange pathway.}
\label{wannier}
\end{figure}
%%%%%%%%%%%%%%%%%%%%%%%%%%%%%%%%%%%%%
The exchange couplings obtained from Eq.\ref{eqn2} and from DFT+\textit{U} calculations are listed in Table.\ref{DFT_L}. The leading interaction $J_\textrm{1}$ runs along the $a$-direction and connects the Mo$^{5+}$ ions into linear spin chains. The mechanism of this interaction is the Mo-O-P-O-Mo superexchange via two equivalent bridges formed by the PO$_4$ tetrahedra. A similar interaction mechanism is known for (VO)$_2$P$_2$O$_7$\cite{Garrett1997}, VOPO$_4 \cdot $0.5H$_2$O\cite{Tennant1997}, and other V$^{4+}$ phosphates, where the double PO$_4$ bridges cause magnetic interactions on the order of 100 K. The interaction in KMoOP$_2$O$_7$ is much weaker, probably due to the lateral displacement of the octahedra, which is known to be unfavorable for superexchange \cite{Roca1998} and indeed restricts the overlap of oxygen $p$-orbitals along the edge of the tetrahedron (Fig.\ref{wannier}). The main interchain interactions, $J_\textrm{2}$ and $J_\textrm{2}^{'}$, are weakly ferromagnetic and zigzag in nature. Residual and non-frustrated coupling, $J_\textrm{b}$, connects the chains along the $b$ direction and is similarly ferromagnetic.

\begin{table}
\centering
\caption{Exchange couplings in KMoOP$_2$O$_7$. The values of $J_\textrm{ij}^\textrm{AFM}$ and $J_\textrm{ij}^\textrm{FM}$ are derived from the Kugel-Khomskii model, \ref{eqn2}, and show the relative contributions of different superexchange mechanisms. Total exchange couplings, $J_\textrm{ij}$, are obtained by the DFT+\textit{U} mapping analysis and may be different from $J_\textrm{ij}^\textrm{AFM}$ and $J_\textrm{ij}^\textrm{FM}$.}
   \renewcommand{\arraystretch}{1.1} % 
    \setlength{\tabcolsep}{8.4pt}
\begin{tabular}{c c c c c}
\hline
\hline
 & $d_\textrm{Mo-Mo}$ (\AA)   & $J_\textrm{ij}^\textrm{AFM}$ (K) & $J_\textrm{ij}^\textrm{FM}$ (K) & $J_\textrm{ij}$ (K)  \\ \hline
$J_\textrm{2}$  & 5.085(1) & 71     & $-$8    & 34   \\
$J_\textrm{1}$  & 5.360(5) & 0      & $-$1.6  & $-$1.4 \\
$J_\textrm{1}^{'}$ & 5.401(5) & 0      & $-$1.4  & $-$1.8 \\
$J_\textrm{b}$  & 6.180(1) & 0      & $-$0.3  & $-$0.4 \\ 
\hline
\hline
\end{tabular}
\label{DFT_L}
\end{table}
%%%%%%%%%%%%%%%%%%%%%%%%%%%%%%%%%%%%%
\begin{figure}
\centering
\includegraphics[width = \columnwidth,keepaspectratio]{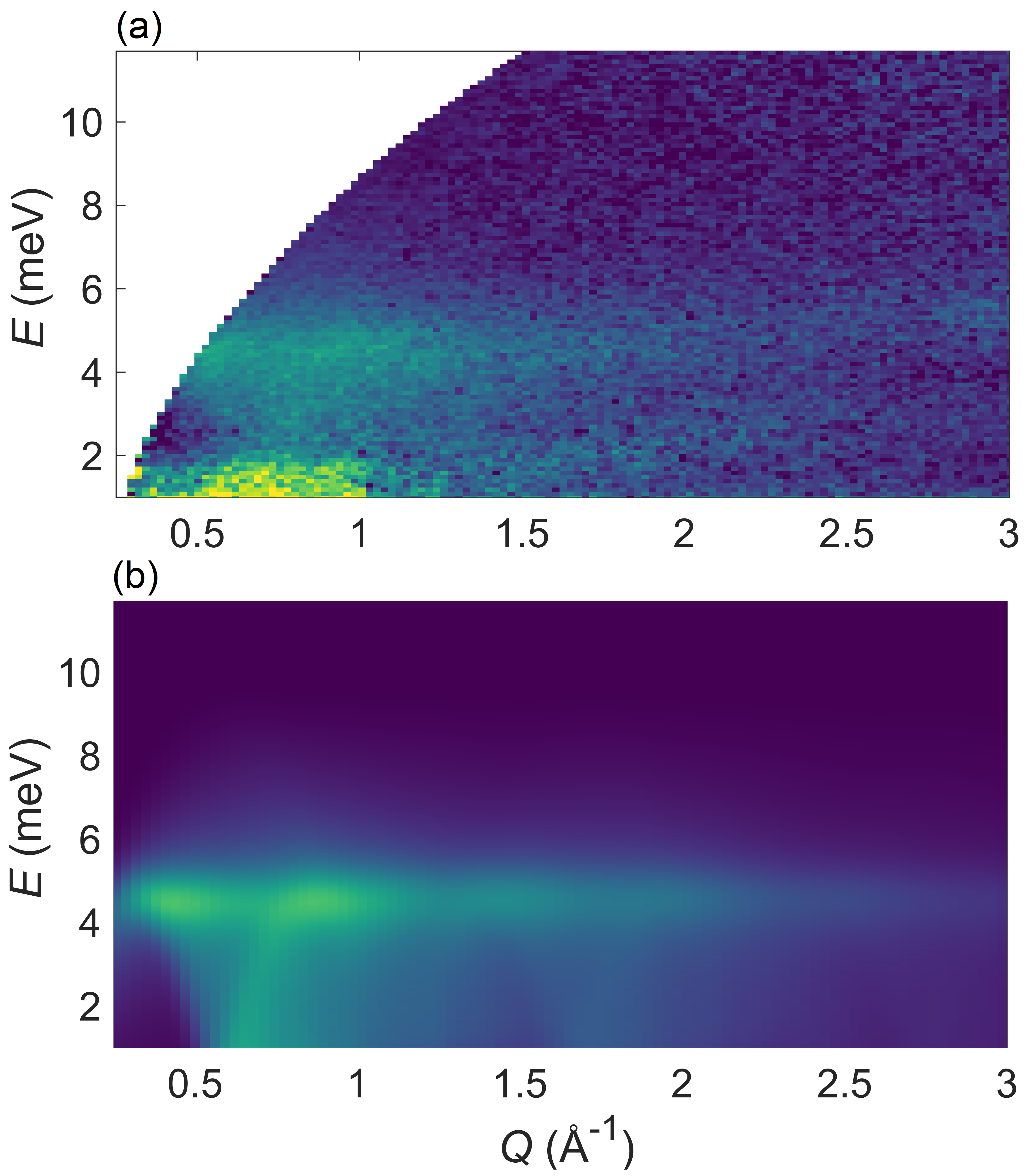}\caption{(a) Phonon subtracted experimental dynamical structure factor, $S(Q,E)$, measured at 1.8 K using an incident energy of $E_\textrm{i} = 14.2$ meV on the IN5 spectrometer. The spurious intensity below 2.5 meV is associated with scattering from the sample environment. (b) Powder-averaged and $E$-resolution convoluted $S\textrm{(}Q,E\textrm{)}^{\textrm{calc}}_\textrm{2+4}$ calculated for $J_\textrm{1} = 34.4(1)$ K and $J_\textrm{2} = - 1.8(1)$ K.}
\label{INS}
\end{figure}
%%%%%%%%%%%%%%%%%%%%%%%%%%%%%%%%%%%%%
%%%%%%%%%%%%%%%%%%%%%%%%%%%%%%%%%%%%%
\begin{figure*}[ht]
\centering
\includegraphics[width = 0.85\textwidth]{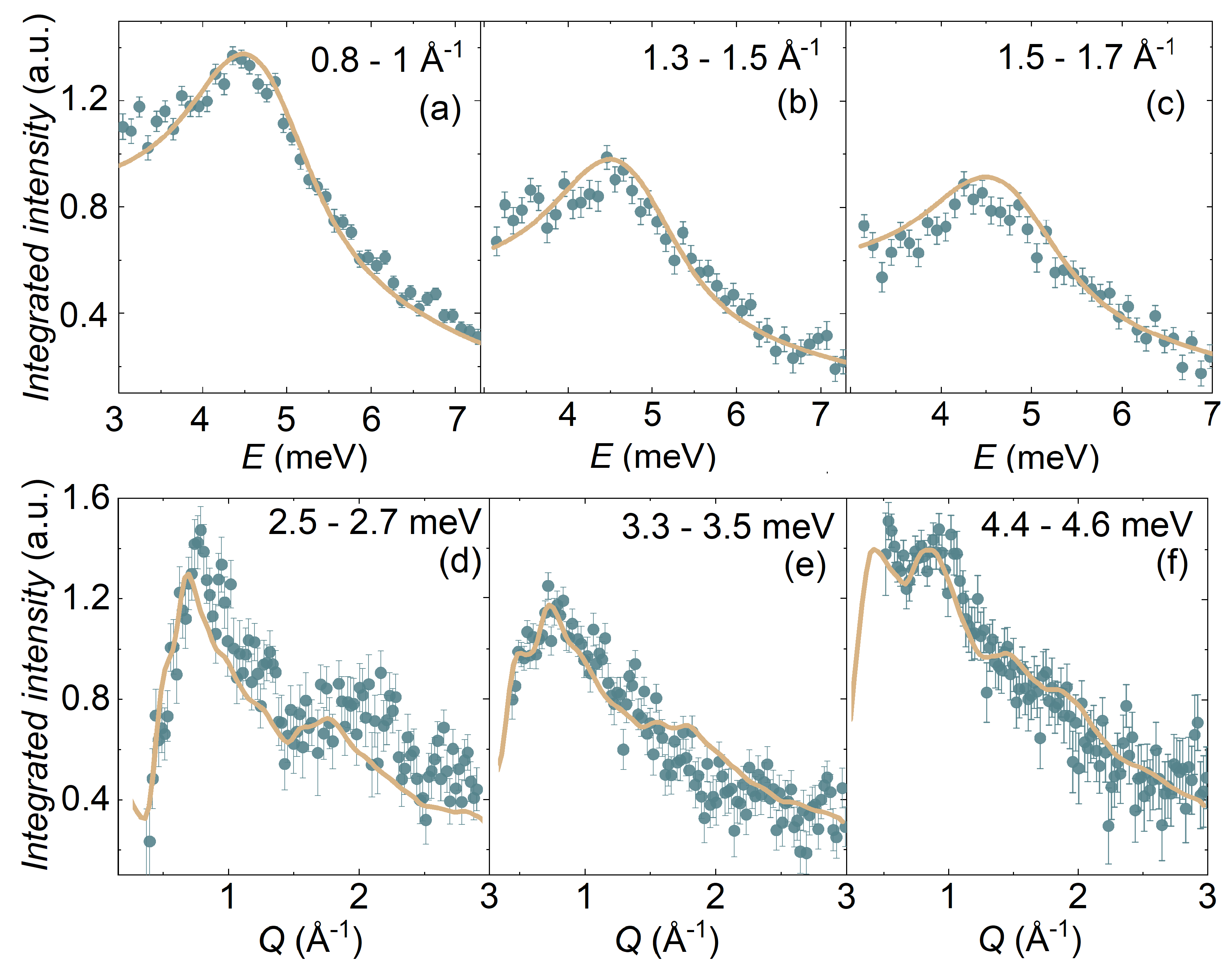}\caption{(a) (a-c)     $\Delta E$- and (d-f) $Q$-integrated cuts (closed circles) fitted to the $S\textrm{(}Q,E\textrm{)}^{\textrm{calc}}_\textrm{2+4}$ calculated as described in the text. The resulting model yields $J_\textrm{2} = 34.4(1)$ K and $J_\textrm{1} = -1.8(2)$ K.}
\label{cuts}
\end{figure*}
%%%%%%%%%%%%%%%%%%%%%%%%%%%%%%%%%%%%%
\subsection{Inelastic Neutron Scattering}
\label{sec:res-INS}
\noindent Having determined the source of the observed one-dimensional magnetic correlations, we now return to the magnetic Hamiltonian of the system. The experimental dynamical structure factor, $S\text{(}Q,E=\hbar \omega \text{)}$, measured at 1.8 K, is shown in Fig.S6(a). At first glance, spurious scattering arising from the sample environment is observed below 2.5 meV at low angles. To correct for this background, we subtract the dynamical structure factor of the diamagnetic analogue (Fig.S6(c)), KNbOP$_2$O$_7$, from the experimental data of KMoOP$_2$O$_7$. This is justified by the similar phonon spectra of the materials. Here, a scaling factor of 1.1 was used to provide the cleanest subtraction. As seen in the resulting spectrum (Fig.\ref{INS}(a)), however, remnants of this spurious scattering, alongside a phonon branch centered about 5.5 meV and peaked around 3 \AA\textsuperscript{-1}, cannot be fully corrected for. As such, our analysis is confined within the $E>2.5$ meV and $Q<3$ \AA\textsuperscript{-1} regions of the spectrum which are least contaminated by these features.

When considering the background subtracted $S\text{(}Q,E\text{)}$ (Fig.\ref{INS}(a)), an accumulation of inelastic spectral weight in a broad feature, peaked about 4.5 meV (Fig.\ref{cuts}(a)), is clearly observed. Alongside the one-dimensional dynamics inferred from the analysis thus far, the sharp onset of magnetic scattering above 0.5 \AA\textsuperscript{-1} (Fig.\ref{cuts}(b-c)) could be associated with powder averaged one-dimensional excitations. Indeed, to a first approximation, the leading exchange parameter, $J_\textrm{1} = 3.02(1)$ meV, extracted from the exact diagonalization fits to $\chi_\textrm{m} \textrm{(}T\textrm{)}$ and $C_\textrm{p, mag} \textrm{(}T\textrm{)}$, coincides with the expected upper limit of the lower bound of the two-spinon continuum, $\pi J/2 \simeq 4.7$ meV.   

To quantify the leading magnetic exchanges in KMoOP$_2$O$_7$, a random phase approximation (RPA)-style approach\cite{Kohno2007} using the dynamical structure factor of the two- and four-spinon continuum, $S\textrm{(}Q,E\textrm{)}^{\textrm{calc}}_\textrm{2+4}$ \cite{Caux2006}, was applied for the Hamiltonian

\begin{equation}
    \mathcal{H} = J_\textrm{2} \sum_{n,n+1}^N S_\textrm{n} \cdot S_\textrm{n+1} + J_\textrm{1} \sum_{n,n+2}^N S_\textrm{n} \cdot S_\textrm{n+2}.
\end{equation}

To optimize the model, four $\Delta E$ cuts, integrated over $\delta Q = 0.1$ \AA \textsuperscript{-1} and centered around 0.9, 1.1, 1.4, and 1.6 \AA \textsuperscript{-1}, were used. The calculated model, powder-averaged and convoluted to the $E$-dependent resolution, was then fit to the experimental data using a particle swarm optimization algorithm. While the resulting model parameters were consistently within the range of 2.93 to 2.96 meV and 0.12 to 0.16 meV for $J_\textrm{1}$ and $\lvert J_\textrm{2} \rvert$, respectively, the sign of the latter was undetermined by the fitting. By considering the effect of the sign of $J_\textrm{1}$ on multiple $\Delta Q$ cuts, integrated over $\delta E = 0.2$ meV, however, the best fit model was found for $J_\textrm{2} = 34.4(1)$~K (2.95(1)~meV) and $J_\textrm{1} = -1.8(2)$~K ($-$0.16(2)~meV) (Fig.\ref{cuts}(a-f)). Indeed, although the redistribution of spectral weight resulting from $J_\textrm{1}$ is expected to be most pronounced within contaminated regions of $S\textrm{(}Q,E\textrm{)}$, a characteristic narrowing of the peaks in $\Delta Q$ was consistent among solutions with ferromagnetic $J_\textrm{1}$ (Fig. S7). Furthermore, the resulting model is consistent with both models obtained through the DFT+\textit{U} calculations (Table.\ref{DFT_L}) and the exact diagonalization fits to $\chi_\textrm{m} \textrm{(}T\textrm{)}$ and $C_\textrm{p, mag} \textrm{(}T\textrm{)}$. Extracting a more conclusive estimate of $J_\textrm{1}$ will require analysis of a dynamical structure factor measured for a single crystalline sample. Regardless of the sign of $J_\textrm{1}$, however, the resulting model is widely consistent with KMoOP$_2$O$_7$ falling within the one-dimensional limit of the HAF phase diagram with $\lvert J_\textrm{1} \rvert / J_\textrm{2} <$ 0.05. Finally, it should be noted that uniform Dzyaloshinskii–Moriya (DM) interactions are allowed along the chain direction. Whilst it cannot be conclusively disregarded, no signs of incommensurate magnetic order, which is expected even for a small DM interaction\cite{dm_exp,dm_THEORY}, could be observed in any data set.
\section{Conclusion}
\label{sec:conc}
\noindent
In summary, we have presented a comprehensive investigation of the structural and magnetic properties of the one-dimensional $S=1/2$ Mo\textsuperscript{5+} system, KMoOP$_2$O$_7$. Using powder neutron diffraction, the previously published structure of KMoOP$_2$O$_7$ has been verified down to 50 mK. Analysis of magnetic susceptibility data revealed one-dimensional magnetic correlations that are consistent with the observed magnetic specific heat. While a broad feature is observed at $T_\textrm{N} = 0.54$ K in the temperature dependence of the specific heat, no magnetic Bragg scattering is observed down to 0.05 K, which we ascribe to the strong one-dimensional character of the correlations in KMoOP$_2$O$_7$. This result is confirmed by our \textit{ab-initio} calculations which reveal an active $d_\textrm{xy}$ magnetic orbital, resulting from distortions to the Mo\textsuperscript{5+}-containing octahedra, favouring superexchange along the chain direction. We finally determine the spin magnetic Hamiltonian through analysis of powder inelastic neutron scattering data resulting in a model that is broadly consistent with all experimental results. We thus propose KMoOP$_2$O$_7$ as a near ideal realization of the $S = 1/2$ Heisenberg one-dimensional chain antiferromagnet model.  

When considering other $S=1/2$ Heisenberg quasi-one-dimensional antiferromagnet materials that are amenable to frustration, the magnetic ground state of KMoOP$_2$O$_7$ is reminiscent to that seen in other materials falling on the one-dimensional limit such as in CuSe\textsubscript{2}O\textsubscript{5}\cite{Janson2009} and Sr\textsubscript{2}CuO\textsubscript{3}\cite{Kojima1997,Rosner1997}. The one-dimensional character of the magnetic correlations in KMoOP$_2$O$_7$ also highlights the significance of the crystalline electric field in determining the effective magnetic ground state. It could then be interesting to examine the role of the monovalent ion site, $A$, in $A$MoOP$_2$O$_7$, in tuning the energy scales in the system, especially given the slight structural variations for $A = $ Na\cite{Ledain1996}, Cs\cite{Guesdon1994}. Investigating the effect of the superexchange mediating ion on the magnetic properties of the system could also prove fruitful by studying KMoV$_2$O$_8$ as the analogous compounds, $AM$P$_2$O$_8$ ($A = $ Tl, K, Rb, Cs, $M = $ Nb, Ta), exist\cite{Paidi2015}. Finally, the roles of the spin and orbital degrees of freedom could be explored in the $5d$ $S = 1/2$ W\textsuperscript{5+} analogue, KWOP$_2$O$_7$\cite{Leclaire2001}.

\section{Acknowledgements}
\noindent Provision of a PhD studentship to A.H.A. by the University of Birmingham and the Science and Technology Facilities Council (STFC) is gratefully acknowledged. The authors are also grateful to the STFC for access to neutron beamtime at ISIS, to Dr. Pascal Manuel (WISH), and thank Dr. Gavin Stenning for aiding with magnetic susceptibility and specific heat measurements at the Materials Characterization Laboratory at ISIS. The Institut Laue-Langevin (ILL) is acknowledged for beam time allocation under the experiment codes 5-31-2822\cite{D20} and EASY-844\cite{IN5} for the D20 and IN5 experiments, respectively. 
\bibliography{References}{}

%merlin.mbs apsrev4-1.bst 2010-07-25 4.21a (PWD, AO, DPC) hacked
%Control: key (0)
%Control: author (8) initials jnrlst
%Control: editor formatted (1) identically to author
%Control: production of article title (-1) disabled
%Control: page (0) single
%Control: year (1) truncated
%Control: production of eprint (0) enabled
\begin{thebibliography}{64}%
\makeatletter
\providecommand \@ifxundefined [1]{%
 \@ifx{#1\undefined}
}%
\providecommand \@ifnum [1]{%
 \ifnum #1\expandafter \@firstoftwo
 \else \expandafter \@secondoftwo
 \fi
}%
\providecommand \@ifx [1]{%
 \ifx #1\expandafter \@firstoftwo
 \else \expandafter \@secondoftwo
 \fi
}%
\providecommand \natexlab [1]{#1}%
\providecommand \enquote  [1]{``#1''}%
\providecommand \bibnamefont  [1]{#1}%
\providecommand \bibfnamefont [1]{#1}%
\providecommand \citenamefont [1]{#1}%
\providecommand \href@noop [0]{\@secondoftwo}%
\providecommand \href [0]{\begingroup \@sanitize@url \@href}%
\providecommand \@href[1]{\@@startlink{#1}\@@href}%
\providecommand \@@href[1]{\endgroup#1\@@endlink}%
\providecommand \@sanitize@url [0]{\catcode `\\12\catcode `\$12\catcode
  `\&12\catcode `\#12\catcode `\^12\catcode `\_12\catcode `\%12\relax}%
\providecommand \@@startlink[1]{}%
\providecommand \@@endlink[0]{}%
\providecommand \url  [0]{\begingroup\@sanitize@url \@url }%
\providecommand \@url [1]{\endgroup\@href {#1}{\urlprefix }}%
\providecommand \urlprefix  [0]{URL }%
\providecommand \Eprint [0]{\href }%
\providecommand \doibase [0]{http://dx.doi.org/}%
\providecommand \selectlanguage [0]{\@gobble}%
\providecommand \bibinfo  [0]{\@secondoftwo}%
\providecommand \bibfield  [0]{\@secondoftwo}%
\providecommand \translation [1]{[#1]}%
\providecommand \BibitemOpen [0]{}%
\providecommand \bibitemStop [0]{}%
\providecommand \bibitemNoStop [0]{.\EOS\space}%
\providecommand \EOS [0]{\spacefactor3000\relax}%
\providecommand \BibitemShut  [1]{\csname bibitem#1\endcsname}%
\let\auto@bib@innerbib\@empty
%</preamble>
\bibitem [{\citenamefont {Nayak}\ \emph {et~al.}(2008)\citenamefont {Nayak},
  \citenamefont {Simon}, \citenamefont {Stern}, \citenamefont {Freedman},\ and\
  \citenamefont {Sarma}}]{Nayak2008}%
  \BibitemOpen
  \bibfield  {author} {\bibinfo {author} {\bibfnamefont {C.}~\bibnamefont
  {Nayak}}, \bibinfo {author} {\bibfnamefont {S.~H.}\ \bibnamefont {Simon}},
  \bibinfo {author} {\bibfnamefont {A.}~\bibnamefont {Stern}}, \bibinfo
  {author} {\bibfnamefont {M.}~\bibnamefont {Freedman}}, \ and\ \bibinfo
  {author} {\bibfnamefont {S.~D.}\ \bibnamefont {Sarma}},\ }\href
  {https://link.aps.org/doi/10.1103/RevModPhys.80.1083} {\bibfield  {journal}
  {\bibinfo  {journal} {Reviews of Modern Physics}\ }\textbf {\bibinfo {volume}
  {80}},\ \bibinfo {pages} {1083} (\bibinfo {year} {2008})}\BibitemShut
  {NoStop}%
\bibitem [{\citenamefont {Kitaev}(2003)}]{Kitaev2003}%
  \BibitemOpen
  \bibfield  {author} {\bibinfo {author} {\bibfnamefont {A.}~\bibnamefont
  {Kitaev}},\ }\href
  {https://www.sciencedirect.com/science/article/pii/S0003491602000180}
  {\bibfield  {journal} {\bibinfo  {journal} {Annals of Physics}\ }\textbf
  {\bibinfo {volume} {303}},\ \bibinfo {pages} {2} (\bibinfo {year}
  {2003})}\BibitemShut {NoStop}%
\bibitem [{\citenamefont {Lahtinen}\ and\ \citenamefont
  {Pachos}(2017)}]{Lahtinen2017}%
  \BibitemOpen
  \bibfield  {author} {\bibinfo {author} {\bibfnamefont {V.}~\bibnamefont
  {Lahtinen}}\ and\ \bibinfo {author} {\bibfnamefont {J.}~\bibnamefont
  {Pachos}},\ }\href {\doibase 10.21468/SciPostPhys.3.3.021} {\bibfield
  {journal} {\bibinfo  {journal} {SciPost Physics}\ }\textbf {\bibinfo {volume}
  {3}},\ \bibinfo {pages} {021} (\bibinfo {year} {2017})}\BibitemShut {NoStop}%
\bibitem [{\citenamefont {Broholm}\ \emph {et~al.}(2020)\citenamefont
  {Broholm}, \citenamefont {Cava}, \citenamefont {Kivelson}, \citenamefont
  {Nocera}, \citenamefont {Norman},\ and\ \citenamefont
  {Senthil}}]{Broholm2020}%
  \BibitemOpen
  \bibfield  {author} {\bibinfo {author} {\bibfnamefont {C.}~\bibnamefont
  {Broholm}}, \bibinfo {author} {\bibfnamefont {R.~J.}\ \bibnamefont {Cava}},
  \bibinfo {author} {\bibfnamefont {S.~A.}\ \bibnamefont {Kivelson}}, \bibinfo
  {author} {\bibfnamefont {D.~G.}\ \bibnamefont {Nocera}}, \bibinfo {author}
  {\bibfnamefont {M.~R.}\ \bibnamefont {Norman}}, \ and\ \bibinfo {author}
  {\bibfnamefont {T.}~\bibnamefont {Senthil}},\ }\href
  {https://doi.org/10.1126/science.aay0668} {\bibfield  {journal} {\bibinfo
  {journal} {Science}\ }\textbf {\bibinfo {volume} {367}},\ \bibinfo {pages}
  {eaay0668} (\bibinfo {year} {2020})}\BibitemShut {NoStop}%
\bibitem [{\citenamefont {Savary}\ and\ \citenamefont
  {Balents}(2016)}]{Savary2016}%
  \BibitemOpen
  \bibfield  {author} {\bibinfo {author} {\bibfnamefont {L.}~\bibnamefont
  {Savary}}\ and\ \bibinfo {author} {\bibfnamefont {L.}~\bibnamefont
  {Balents}},\ }\href {\doibase 10.1088/0034-4885/80/1/016502} {\bibfield
  {journal} {\bibinfo  {journal} {Reports on Progress in Physics}\ }\textbf
  {\bibinfo {volume} {80}},\ \bibinfo {pages} {016502} (\bibinfo {year}
  {2016})}\BibitemShut {NoStop}%
\bibitem [{\citenamefont {Clark}\ and\ \citenamefont
  {Abdeldaim}(2021)}]{Clark2021}%
  \BibitemOpen
  \bibfield  {author} {\bibinfo {author} {\bibfnamefont {L.}~\bibnamefont
  {Clark}}\ and\ \bibinfo {author} {\bibfnamefont {A.~H.}\ \bibnamefont
  {Abdeldaim}},\ }\href {https://doi.org/10.1146/annurev-matsci-080819-011453}
  {\bibfield  {journal} {\bibinfo  {journal} {Annual Review of Materials
  Research}\ }\textbf {\bibinfo {volume} {51}},\ \bibinfo {pages} {495}
  (\bibinfo {year} {2021})}\BibitemShut {NoStop}%
\bibitem [{\citenamefont {Castelnovo}\ \emph {et~al.}(2012)\citenamefont
  {Castelnovo}, \citenamefont {Moessner},\ and\ \citenamefont
  {Sondhi}}]{Castelnovo2012}%
  \BibitemOpen
  \bibfield  {author} {\bibinfo {author} {\bibfnamefont {C.}~\bibnamefont
  {Castelnovo}}, \bibinfo {author} {\bibfnamefont {R.}~\bibnamefont
  {Moessner}}, \ and\ \bibinfo {author} {\bibfnamefont {S.~L.}\ \bibnamefont
  {Sondhi}},\ }\href {\doibase 10.1146/annurev-conmatphys-020911-125058}
  {\bibfield  {journal} {\bibinfo  {journal} {Annual Review of Condensed Matter
  Physics}\ }\textbf {\bibinfo {volume} {3}},\ \bibinfo {pages} {35} (\bibinfo
  {year} {2012})}\BibitemShut {NoStop}%
\bibitem [{\citenamefont {Hermanns}\ \emph {et~al.}(2018)\citenamefont
  {Hermanns}, \citenamefont {Kimchi},\ and\ \citenamefont
  {Knolle}}]{Hermanns2018}%
  \BibitemOpen
  \bibfield  {author} {\bibinfo {author} {\bibfnamefont {M.}~\bibnamefont
  {Hermanns}}, \bibinfo {author} {\bibfnamefont {I.}~\bibnamefont {Kimchi}}, \
  and\ \bibinfo {author} {\bibfnamefont {J.}~\bibnamefont {Knolle}},\ }\href
  {\doibase 10.1146/annurev-conmatphys-033117-053934} {\bibfield  {journal}
  {\bibinfo  {journal} {Annual Review of Condensed Matter Physics}\ }\textbf
  {\bibinfo {volume} {9}},\ \bibinfo {pages} {17} (\bibinfo {year}
  {2018})}\BibitemShut {NoStop}%
\bibitem [{\citenamefont {Feldman}\ and\ \citenamefont
  {Halperin}(2021)}]{Feldman2021}%
  \BibitemOpen
  \bibfield  {author} {\bibinfo {author} {\bibfnamefont {D.~E.}\ \bibnamefont
  {Feldman}}\ and\ \bibinfo {author} {\bibfnamefont {B.~I.}\ \bibnamefont
  {Halperin}},\ }\href {http://dx.doi.org/10.1088/1361-6633/ac03aa} {\bibfield
  {journal} {\bibinfo  {journal} {Reports on Progress in Physics}\ }\textbf
  {\bibinfo {volume} {84}},\ \bibinfo {pages} {076501} (\bibinfo {year}
  {2021})}\BibitemShut {NoStop}%
\bibitem [{\citenamefont {Bethe}(1931)}]{Bethe1931}%
  \BibitemOpen
  \bibfield  {author} {\bibinfo {author} {\bibfnamefont {H.}~\bibnamefont
  {Bethe}},\ }\href {https://doi.org/10.1007/BF01341708} {\bibfield  {journal}
  {\bibinfo  {journal} {Zeitschrift für Physik}\ }\textbf {\bibinfo {volume}
  {71}},\ \bibinfo {pages} {205} (\bibinfo {year} {1931})}\BibitemShut
  {NoStop}%
\bibitem [{\citenamefont {Hulth\'en}(1938)}]{Hulthen1938}%
  \BibitemOpen
  \bibfield  {author} {\bibinfo {author} {\bibfnamefont {L.}~\bibnamefont
  {Hulth\'en}},\ }\href@noop {} {\enquote {\bibinfo {title} {Uber das
  austauschproblem eines kristalles.}}\ } (\bibinfo {year} {1938})\BibitemShut
  {NoStop}%
\bibitem [{\citenamefont {Tennant}\ \emph {et~al.}(1993)\citenamefont
  {Tennant}, \citenamefont {Perring}, \citenamefont {Cowley},\ and\
  \citenamefont {Nagler}}]{Tennant1993}%
  \BibitemOpen
  \bibfield  {author} {\bibinfo {author} {\bibfnamefont {D.~A.}\ \bibnamefont
  {Tennant}}, \bibinfo {author} {\bibfnamefont {T.~G.}\ \bibnamefont
  {Perring}}, \bibinfo {author} {\bibfnamefont {R.~A.}\ \bibnamefont {Cowley}},
  \ and\ \bibinfo {author} {\bibfnamefont {S.~E.}\ \bibnamefont {Nagler}},\
  }\href {\doibase 10.1103/PhysRevLett.70.4003} {\bibfield  {journal} {\bibinfo
   {journal} {Physical Review Letters}\ }\textbf {\bibinfo {volume} {70}},\
  \bibinfo {pages} {4003} (\bibinfo {year} {1993})}\BibitemShut {NoStop}%
\bibitem [{\citenamefont {Mourigal}\ \emph {et~al.}(2013)\citenamefont
  {Mourigal}, \citenamefont {Enderle}, \citenamefont {Klöpperpieper},
  \citenamefont {Caux}, \citenamefont {Stunault},\ and\ \citenamefont
  {Rønnow}}]{Mourigal2013}%
  \BibitemOpen
  \bibfield  {author} {\bibinfo {author} {\bibfnamefont {M.}~\bibnamefont
  {Mourigal}}, \bibinfo {author} {\bibfnamefont {M.}~\bibnamefont {Enderle}},
  \bibinfo {author} {\bibfnamefont {A.}~\bibnamefont {Klöpperpieper}},
  \bibinfo {author} {\bibfnamefont {J.-S.}\ \bibnamefont {Caux}}, \bibinfo
  {author} {\bibfnamefont {A.}~\bibnamefont {Stunault}}, \ and\ \bibinfo
  {author} {\bibfnamefont {H.~M.}\ \bibnamefont {Rønnow}},\ }\href {\doibase
  10.1038/nphys2652} {\bibfield  {journal} {\bibinfo  {journal} {Nature
  Physics}\ }\textbf {\bibinfo {volume} {9}},\ \bibinfo {pages} {435} (\bibinfo
  {year} {2013})}\BibitemShut {NoStop}%
\bibitem [{\citenamefont {Coldea}\ \emph {et~al.}(2010)\citenamefont {Coldea},
  \citenamefont {Tennant}, \citenamefont {Wheeler}, \citenamefont {Wawrzynska},
  \citenamefont {Prabhakaran}, \citenamefont {Telling}, \citenamefont
  {Habicht}, \citenamefont {Smeibidl},\ and\ \citenamefont {Kiefer}}]{Coldea}%
  \BibitemOpen
  \bibfield  {author} {\bibinfo {author} {\bibfnamefont {R.}~\bibnamefont
  {Coldea}}, \bibinfo {author} {\bibfnamefont {D.~A.}\ \bibnamefont {Tennant}},
  \bibinfo {author} {\bibfnamefont {E.~M.}\ \bibnamefont {Wheeler}}, \bibinfo
  {author} {\bibfnamefont {E.}~\bibnamefont {Wawrzynska}}, \bibinfo {author}
  {\bibfnamefont {D.}~\bibnamefont {Prabhakaran}}, \bibinfo {author}
  {\bibfnamefont {M.}~\bibnamefont {Telling}}, \bibinfo {author} {\bibfnamefont
  {K.}~\bibnamefont {Habicht}}, \bibinfo {author} {\bibfnamefont
  {P.}~\bibnamefont {Smeibidl}}, \ and\ \bibinfo {author} {\bibfnamefont
  {K.}~\bibnamefont {Kiefer}},\ }\href {\doibase 10.1126/science.1180085}
  {\bibfield  {journal} {\bibinfo  {journal} {Science}\ }\textbf {\bibinfo
  {volume} {327}},\ \bibinfo {pages} {177} (\bibinfo {year}
  {2010})}\BibitemShut {NoStop}%
\bibitem [{\citenamefont {Wang}\ \emph {et~al.}(2015)\citenamefont {Wang},
  \citenamefont {Schmidt}, \citenamefont {Bera}, \citenamefont {Islam},
  \citenamefont {Lake}, \citenamefont {Loidl},\ and\ \citenamefont
  {Deisenhofer}}]{Wang2015}%
  \BibitemOpen
  \bibfield  {author} {\bibinfo {author} {\bibfnamefont {Z.}~\bibnamefont
  {Wang}}, \bibinfo {author} {\bibfnamefont {M.}~\bibnamefont {Schmidt}},
  \bibinfo {author} {\bibfnamefont {A.~K.}\ \bibnamefont {Bera}}, \bibinfo
  {author} {\bibfnamefont {A.~T. M.~N.}\ \bibnamefont {Islam}}, \bibinfo
  {author} {\bibfnamefont {B.}~\bibnamefont {Lake}}, \bibinfo {author}
  {\bibfnamefont {A.}~\bibnamefont {Loidl}}, \ and\ \bibinfo {author}
  {\bibfnamefont {J.}~\bibnamefont {Deisenhofer}},\ }\href {\doibase
  10.1103/PhysRevB.91.140404} {\bibfield  {journal} {\bibinfo  {journal}
  {Physical Review B}\ }\textbf {\bibinfo {volume} {91}},\ \bibinfo {pages}
  {140404} (\bibinfo {year} {2015})}\BibitemShut {NoStop}%
\bibitem [{\citenamefont {Vasiliev}\ \emph {et~al.}(2018)\citenamefont
  {Vasiliev}, \citenamefont {Volkova}, \citenamefont {Zvereva},\ and\
  \citenamefont {Markina}}]{Vasiliev2018}%
  \BibitemOpen
  \bibfield  {author} {\bibinfo {author} {\bibfnamefont {A.}~\bibnamefont
  {Vasiliev}}, \bibinfo {author} {\bibfnamefont {O.}~\bibnamefont {Volkova}},
  \bibinfo {author} {\bibfnamefont {E.}~\bibnamefont {Zvereva}}, \ and\
  \bibinfo {author} {\bibfnamefont {M.}~\bibnamefont {Markina}},\ }\href
  {\doibase 10.1038/s41535-018-0090-7} {\bibfield  {journal} {\bibinfo
  {journal} {npj Quantum Materials}\ }\textbf {\bibinfo {volume} {3}},\
  \bibinfo {pages} {18} (\bibinfo {year} {2018})}\BibitemShut {NoStop}%
\bibitem [{\citenamefont {Haldane}(1982)}]{Haldane1982}%
  \BibitemOpen
  \bibfield  {author} {\bibinfo {author} {\bibfnamefont {F.~D.~M.}\
  \bibnamefont {Haldane}},\ }\href {\doibase 10.1103/PhysRevB.25.4925}
  {\bibfield  {journal} {\bibinfo  {journal} {Physical Review B}\ }\textbf
  {\bibinfo {volume} {25}},\ \bibinfo {pages} {4925} (\bibinfo {year}
  {1982})}\BibitemShut {NoStop}%
\bibitem [{\citenamefont {Okamoto}\ and\ \citenamefont
  {Nomura}(1992)}]{Okamoto1992}%
  \BibitemOpen
  \bibfield  {author} {\bibinfo {author} {\bibfnamefont {K.}~\bibnamefont
  {Okamoto}}\ and\ \bibinfo {author} {\bibfnamefont {K.}~\bibnamefont
  {Nomura}},\ }\href {\doibase https://doi.org/10.1016/0375-9601(92)90823-5}
  {\bibfield  {journal} {\bibinfo  {journal} {Physics Letters A}\ }\textbf
  {\bibinfo {volume} {169}},\ \bibinfo {pages} {433} (\bibinfo {year}
  {1992})}\BibitemShut {NoStop}%
\bibitem [{\citenamefont {Eggert}(1996)}]{Eggert1996}%
  \BibitemOpen
  \bibfield  {author} {\bibinfo {author} {\bibfnamefont {S.}~\bibnamefont
  {Eggert}},\ }\href {\doibase 10.1103/PhysRevB.54.R9612} {\bibfield  {journal}
  {\bibinfo  {journal} {Physical Review B}\ }\textbf {\bibinfo {volume} {54}},\
  \bibinfo {pages} {R9612} (\bibinfo {year} {1996})}\BibitemShut {NoStop}%
\bibitem [{\citenamefont {Enderle}\ \emph {et~al.}(2005)\citenamefont
  {Enderle}, \citenamefont {Mukherjee}, \citenamefont {Fåk}, \citenamefont
  {Kremer}, \citenamefont {Broto}, \citenamefont {Rosner}, \citenamefont
  {Drechsler}, \citenamefont {Richter}, \citenamefont {Malek}, \citenamefont
  {Prokofiev}, \citenamefont {Assmus}, \citenamefont {Pujol}, \citenamefont
  {Raggazzoni}, \citenamefont {Rakoto}, \citenamefont {Rheinstädter},\ and\
  \citenamefont {Rønnow}}]{Enderle2005}%
  \BibitemOpen
  \bibfield  {author} {\bibinfo {author} {\bibfnamefont {M.}~\bibnamefont
  {Enderle}}, \bibinfo {author} {\bibfnamefont {C.}~\bibnamefont {Mukherjee}},
  \bibinfo {author} {\bibfnamefont {B.}~\bibnamefont {Fåk}}, \bibinfo {author}
  {\bibfnamefont {R.~K.}\ \bibnamefont {Kremer}}, \bibinfo {author}
  {\bibfnamefont {J.-M.}\ \bibnamefont {Broto}}, \bibinfo {author}
  {\bibfnamefont {H.}~\bibnamefont {Rosner}}, \bibinfo {author} {\bibfnamefont
  {S.-L.}\ \bibnamefont {Drechsler}}, \bibinfo {author} {\bibfnamefont
  {J.}~\bibnamefont {Richter}}, \bibinfo {author} {\bibfnamefont
  {J.}~\bibnamefont {Malek}}, \bibinfo {author} {\bibfnamefont
  {A.}~\bibnamefont {Prokofiev}}, \bibinfo {author} {\bibfnamefont
  {W.}~\bibnamefont {Assmus}}, \bibinfo {author} {\bibfnamefont
  {S.}~\bibnamefont {Pujol}}, \bibinfo {author} {\bibfnamefont {J.-L.}\
  \bibnamefont {Raggazzoni}}, \bibinfo {author} {\bibfnamefont
  {H.}~\bibnamefont {Rakoto}}, \bibinfo {author} {\bibfnamefont
  {M.}~\bibnamefont {Rheinstädter}}, \ and\ \bibinfo {author} {\bibfnamefont
  {H.~M.}\ \bibnamefont {Rønnow}},\ }\href {\doibase
  10.1209/epl/i2004-10484-x} {\bibfield  {journal} {\bibinfo  {journal}
  {Europhysics Letters (EPL)}\ }\textbf {\bibinfo {volume} {70}},\ \bibinfo
  {pages} {237} (\bibinfo {year} {2005})}\BibitemShut {NoStop}%
\bibitem [{\citenamefont {Dutton}\ \emph {et~al.}(2012)\citenamefont {Dutton},
  \citenamefont {Kumar}, \citenamefont {Mourigal}, \citenamefont {Soos},
  \citenamefont {Wen}, \citenamefont {Broholm}, \citenamefont {Andersen},
  \citenamefont {Huang}, \citenamefont {Zbiri}, \citenamefont {Toft-Petersen},\
  and\ \citenamefont {Cava}}]{Dutton2012}%
  \BibitemOpen
  \bibfield  {author} {\bibinfo {author} {\bibfnamefont {S.~E.}\ \bibnamefont
  {Dutton}}, \bibinfo {author} {\bibfnamefont {M.}~\bibnamefont {Kumar}},
  \bibinfo {author} {\bibfnamefont {M.}~\bibnamefont {Mourigal}}, \bibinfo
  {author} {\bibfnamefont {Z.~G.}\ \bibnamefont {Soos}}, \bibinfo {author}
  {\bibfnamefont {J.-J.}\ \bibnamefont {Wen}}, \bibinfo {author} {\bibfnamefont
  {C.~L.}\ \bibnamefont {Broholm}}, \bibinfo {author} {\bibfnamefont {N.~H.}\
  \bibnamefont {Andersen}}, \bibinfo {author} {\bibfnamefont {Q.}~\bibnamefont
  {Huang}}, \bibinfo {author} {\bibfnamefont {M.}~\bibnamefont {Zbiri}},
  \bibinfo {author} {\bibfnamefont {R.}~\bibnamefont {Toft-Petersen}}, \ and\
  \bibinfo {author} {\bibfnamefont {R.~J.}\ \bibnamefont {Cava}},\ }\href
  {https://link.aps.org/doi/10.1103/PhysRevLett.108.187206} {\bibfield
  {journal} {\bibinfo  {journal} {Physical Review Letters}\ }\textbf {\bibinfo
  {volume} {108}},\ \bibinfo {pages} {187206} (\bibinfo {year}
  {2012})}\BibitemShut {NoStop}%
\bibitem [{\citenamefont {Nilsen}\ \emph {et~al.}(2008)\citenamefont {Nilsen},
  \citenamefont {Rønnow}, \citenamefont {Läuchli}, \citenamefont {Fabbiani},
  \citenamefont {Sanchez-Benitez}, \citenamefont {Kamenev},\ and\ \citenamefont
  {Harrison}}]{Nilsen2008}%
  \BibitemOpen
  \bibfield  {author} {\bibinfo {author} {\bibfnamefont {G.~J.}\ \bibnamefont
  {Nilsen}}, \bibinfo {author} {\bibfnamefont {H.~M.}\ \bibnamefont {Rønnow}},
  \bibinfo {author} {\bibfnamefont {A.~M.}\ \bibnamefont {Läuchli}}, \bibinfo
  {author} {\bibfnamefont {F.~P.~A.}\ \bibnamefont {Fabbiani}}, \bibinfo
  {author} {\bibfnamefont {J.}~\bibnamefont {Sanchez-Benitez}}, \bibinfo
  {author} {\bibfnamefont {K.~V.}\ \bibnamefont {Kamenev}}, \ and\ \bibinfo
  {author} {\bibfnamefont {A.}~\bibnamefont {Harrison}},\ }\href {\doibase
  10.1021/cm7023263} {\bibfield  {journal} {\bibinfo  {journal} {Chemistry of
  Materials}\ }\textbf {\bibinfo {volume} {20}},\ \bibinfo {pages} {8}
  (\bibinfo {year} {2008})}\BibitemShut {NoStop}%
\bibitem [{\citenamefont {Kasinathan}\ \emph {et~al.}(2013)\citenamefont
  {Kasinathan}, \citenamefont {Koepernik}, \citenamefont {Janson},
  \citenamefont {Nilsen}, \citenamefont {Piatek}, \citenamefont {Rønnow},\
  and\ \citenamefont {Rosner}}]{Kasinathan2013}%
  \BibitemOpen
  \bibfield  {author} {\bibinfo {author} {\bibfnamefont {D.}~\bibnamefont
  {Kasinathan}}, \bibinfo {author} {\bibfnamefont {K.}~\bibnamefont
  {Koepernik}}, \bibinfo {author} {\bibfnamefont {O.}~\bibnamefont {Janson}},
  \bibinfo {author} {\bibfnamefont {G.~J.}\ \bibnamefont {Nilsen}}, \bibinfo
  {author} {\bibfnamefont {J.~O.}\ \bibnamefont {Piatek}}, \bibinfo {author}
  {\bibfnamefont {H.~M.}\ \bibnamefont {Rønnow}}, \ and\ \bibinfo {author}
  {\bibfnamefont {H.}~\bibnamefont {Rosner}},\ }\href {\doibase
  10.1103/PhysRevB.88.224410} {\bibfield  {journal} {\bibinfo  {journal}
  {Physical Review B}\ }\textbf {\bibinfo {volume} {88}},\ \bibinfo {pages}
  {224410} (\bibinfo {year} {2013})}\BibitemShut {NoStop}%
\bibitem [{\citenamefont {Gueho}\ \emph {et~al.}(1992)\citenamefont {Gueho},
  \citenamefont {Borel}, \citenamefont {Grandin}, \citenamefont {Leclaire},\
  and\ \citenamefont {Raveau}}]{Gueho1992}%
  \BibitemOpen
  \bibfield  {author} {\bibinfo {author} {\bibfnamefont {C.}~\bibnamefont
  {Gueho}}, \bibinfo {author} {\bibfnamefont {M.~M.}\ \bibnamefont {Borel}},
  \bibinfo {author} {\bibfnamefont {A.}~\bibnamefont {Grandin}}, \bibinfo
  {author} {\bibfnamefont {A.}~\bibnamefont {Leclaire}}, \ and\ \bibinfo
  {author} {\bibfnamefont {B.}~\bibnamefont {Raveau}},\ }\href
  {https://doi.org/10.1002/zaac.19926150921} {\bibfield  {journal} {\bibinfo
  {journal} {Zeitschrift für anorganische und allgemeine Chemie}\ }\textbf
  {\bibinfo {volume} {615}},\ \bibinfo {pages} {104} (\bibinfo {year}
  {1992})}\BibitemShut {NoStop}%
\bibitem [{\citenamefont {Canadell}\ \emph {et~al.}(1997)\citenamefont
  {Canadell}, \citenamefont {Provost}, \citenamefont {Guesdon}, \citenamefont
  {Borel},\ and\ \citenamefont {Leclaire}}]{Canadell1997}%
  \BibitemOpen
  \bibfield  {author} {\bibinfo {author} {\bibfnamefont {E.}~\bibnamefont
  {Canadell}}, \bibinfo {author} {\bibfnamefont {J.}~\bibnamefont {Provost}},
  \bibinfo {author} {\bibfnamefont {A.}~\bibnamefont {Guesdon}}, \bibinfo
  {author} {\bibfnamefont {M.~M.}\ \bibnamefont {Borel}}, \ and\ \bibinfo
  {author} {\bibfnamefont {A.}~\bibnamefont {Leclaire}},\ }\href {\doibase
  10.1021/cm960161v} {\bibfield  {journal} {\bibinfo  {journal} {Chemistry of
  Materials}\ }\textbf {\bibinfo {volume} {9}},\ \bibinfo {pages} {68}
  (\bibinfo {year} {1997})}\BibitemShut {NoStop}%
\bibitem [{\citenamefont {Ledain}\ \emph {et~al.}(1996)\citenamefont {Ledain},
  \citenamefont {Leclaire}, \citenamefont {Borel}, \citenamefont {Provost},\
  and\ \citenamefont {Raveau}}]{Ledain1996}%
  \BibitemOpen
  \bibfield  {author} {\bibinfo {author} {\bibfnamefont {S.}~\bibnamefont
  {Ledain}}, \bibinfo {author} {\bibfnamefont {A.}~\bibnamefont {Leclaire}},
  \bibinfo {author} {\bibfnamefont {M.~M.}\ \bibnamefont {Borel}}, \bibinfo
  {author} {\bibfnamefont {J.}~\bibnamefont {Provost}}, \ and\ \bibinfo
  {author} {\bibfnamefont {B.}~\bibnamefont {Raveau}},\ }\href
  {https://www.sciencedirect.com/science/article/pii/S0022459696902029}
  {\bibfield  {journal} {\bibinfo  {journal} {Journal of Solid State
  Chemistry}\ }\textbf {\bibinfo {volume} {124}},\ \bibinfo {pages} {24}
  (\bibinfo {year} {1996})}\BibitemShut {NoStop}%
\bibitem [{\citenamefont {Guesdon}\ \emph {et~al.}(1994)\citenamefont
  {Guesdon}, \citenamefont {Borel}, \citenamefont {Leclaire}, \citenamefont
  {Grandin},\ and\ \citenamefont {Raveau}}]{Guesdon1994}%
  \BibitemOpen
  \bibfield  {author} {\bibinfo {author} {\bibfnamefont {A.}~\bibnamefont
  {Guesdon}}, \bibinfo {author} {\bibfnamefont {M.~M.}\ \bibnamefont {Borel}},
  \bibinfo {author} {\bibfnamefont {A.}~\bibnamefont {Leclaire}}, \bibinfo
  {author} {\bibfnamefont {A.}~\bibnamefont {Grandin}}, \ and\ \bibinfo
  {author} {\bibfnamefont {B.}~\bibnamefont {Raveau}},\ }\href
  {https://www.sciencedirect.com/science/article/pii/S0022459684710073}
  {\bibfield  {journal} {\bibinfo  {journal} {Journal of Solid State
  Chemistry}\ }\textbf {\bibinfo {volume} {108}},\ \bibinfo {pages} {46}
  (\bibinfo {year} {1994})}\BibitemShut {NoStop}%
\bibitem [{\citenamefont {Momma}\ and\ \citenamefont
  {Izumi}(2008)}]{Momma:ko5060}%
  \BibitemOpen
  \bibfield  {author} {\bibinfo {author} {\bibfnamefont {K.}~\bibnamefont
  {Momma}}\ and\ \bibinfo {author} {\bibfnamefont {F.}~\bibnamefont {Izumi}},\
  }\href {https://doi.org/10.1107/S0021889808012016} {\bibfield  {journal}
  {\bibinfo  {journal} {Journal of Applied Crystallography}\ }\textbf {\bibinfo
  {volume} {41}},\ \bibinfo {pages} {653} (\bibinfo {year} {2008})}\BibitemShut
  {NoStop}%
\bibitem [{\citenamefont {Abdeldaim}\ \emph
  {et~al.}(2021{\natexlab{a}})\citenamefont {Abdeldaim}, \citenamefont {Clark},
  \citenamefont {Nilsen},\ and\ \citenamefont {Ritter}}]{D20}%
  \BibitemOpen
  \bibfield  {author} {\bibinfo {author} {\bibfnamefont {A.~H.}\ \bibnamefont
  {Abdeldaim}}, \bibinfo {author} {\bibfnamefont {L.}~\bibnamefont {Clark}},
  \bibinfo {author} {\bibfnamefont {G.~J.}\ \bibnamefont {Nilsen}}, \ and\
  \bibinfo {author} {\bibfnamefont {C.}~\bibnamefont {Ritter}},\ }\href
  {\doibase https://doi.ill.fr/10.5291/ILL-DATA.5-31-2886} {\  (\bibinfo {year}
  {2021}{\natexlab{a}}),\
  https://doi.ill.fr/10.5291/ILL-DATA.5-31-2886}\BibitemShut {NoStop}%
\bibitem [{\citenamefont {Toby}(2001)}]{Toby2001}%
  \BibitemOpen
  \bibfield  {author} {\bibinfo {author} {\bibfnamefont {B.~H.}\ \bibnamefont
  {Toby}},\ }\href {\doibase 10.1107/S0021889801002242} {\bibfield  {journal}
  {\bibinfo  {journal} {Journal of Applied Crystallography}\ }\textbf {\bibinfo
  {volume} {34}},\ \bibinfo {pages} {210} (\bibinfo {year} {2001})}\BibitemShut
  {NoStop}%
\bibitem [{\citenamefont {Qureshi}(2019)}]{Qureshi2019}%
  \BibitemOpen
  \bibfield  {author} {\bibinfo {author} {\bibfnamefont {N.}~\bibnamefont
  {Qureshi}},\ }\href {\doibase 10.1107/S1600576718016084} {\bibfield
  {journal} {\bibinfo  {journal} {Journal of Applied Crystallography}\ }\textbf
  {\bibinfo {volume} {52}},\ \bibinfo {pages} {175} (\bibinfo {year}
  {2019})}\BibitemShut {NoStop}%
\bibitem [{\citenamefont {Ishikawa}\ \emph {et~al.}(2017)\citenamefont
  {Ishikawa}, \citenamefont {Nakamura}, \citenamefont {Yoshida}, \citenamefont
  {Takigawa}, \citenamefont {Babkevich}, \citenamefont {Qureshi}, \citenamefont
  {Rønnow}, \citenamefont {Yajima},\ and\ \citenamefont
  {Hiroi}}]{Ishikawa2017}%
  \BibitemOpen
  \bibfield  {author} {\bibinfo {author} {\bibfnamefont {H.}~\bibnamefont
  {Ishikawa}}, \bibinfo {author} {\bibfnamefont {N.}~\bibnamefont {Nakamura}},
  \bibinfo {author} {\bibfnamefont {M.}~\bibnamefont {Yoshida}}, \bibinfo
  {author} {\bibfnamefont {M.}~\bibnamefont {Takigawa}}, \bibinfo {author}
  {\bibfnamefont {P.}~\bibnamefont {Babkevich}}, \bibinfo {author}
  {\bibfnamefont {N.}~\bibnamefont {Qureshi}}, \bibinfo {author} {\bibfnamefont
  {H.~M.}\ \bibnamefont {Rønnow}}, \bibinfo {author} {\bibfnamefont
  {T.}~\bibnamefont {Yajima}}, \ and\ \bibinfo {author} {\bibfnamefont
  {Z.}~\bibnamefont {Hiroi}},\ }\href
  {https://link.aps.org/doi/10.1103/PhysRevB.95.064408} {\bibfield  {journal}
  {\bibinfo  {journal} {Physical Review B}\ }\textbf {\bibinfo {volume} {95}},\
  \bibinfo {pages} {64408} (\bibinfo {year} {2017})}\BibitemShut {NoStop}%
\bibitem [{\citenamefont {Abdeldaim}\ \emph
  {et~al.}(2021{\natexlab{b}})\citenamefont {Abdeldaim}, \citenamefont {Clark},
  \citenamefont {Nilsen},\ and\ \citenamefont {Ollivier}}]{IN5}%
  \BibitemOpen
  \bibfield  {author} {\bibinfo {author} {\bibfnamefont {A.~H.}\ \bibnamefont
  {Abdeldaim}}, \bibinfo {author} {\bibfnamefont {L.}~\bibnamefont {Clark}},
  \bibinfo {author} {\bibfnamefont {G.~J.}\ \bibnamefont {Nilsen}}, \ and\
  \bibinfo {author} {\bibfnamefont {J.}~\bibnamefont {Ollivier}},\ }\href
  {\doibase https://doi.ill.fr/10.5291/ILL-DATA.EASY-844} {\  (\bibinfo {year}
  {2021}{\natexlab{b}}),\
  https://doi.ill.fr/10.5291/ILL-DATA.EASY-844}\BibitemShut {NoStop}%
\bibitem [{\citenamefont {Caux}\ and\ \citenamefont
  {Hagemans}(2006)}]{Caux2006}%
  \BibitemOpen
  \bibfield  {author} {\bibinfo {author} {\bibfnamefont {J.-S.}\ \bibnamefont
  {Caux}}\ and\ \bibinfo {author} {\bibfnamefont {R.}~\bibnamefont
  {Hagemans}},\ }\href {\doibase 10.1088/1742-5468/2006/12/p12013} {\bibfield
  {journal} {\bibinfo  {journal} {Journal of Statistical Mechanics: Theory and
  Experiment}\ }\textbf {\bibinfo {volume} {2006}},\ \bibinfo {pages} {P12013}
  (\bibinfo {year} {2006})}\BibitemShut {NoStop}%
\bibitem [{\citenamefont {Kohno}\ \emph {et~al.}(2007)\citenamefont {Kohno},
  \citenamefont {Starykh},\ and\ \citenamefont {Balents}}]{Kohno2007}%
  \BibitemOpen
  \bibfield  {author} {\bibinfo {author} {\bibfnamefont {M.}~\bibnamefont
  {Kohno}}, \bibinfo {author} {\bibfnamefont {O.~A.}\ \bibnamefont {Starykh}},
  \ and\ \bibinfo {author} {\bibfnamefont {L.}~\bibnamefont {Balents}},\ }\href
  {\doibase 10.1038/nphys749} {\bibfield  {journal} {\bibinfo  {journal}
  {Nature Physics}\ }\textbf {\bibinfo {volume} {3}},\ \bibinfo {pages} {790}
  (\bibinfo {year} {2007})}\BibitemShut {NoStop}%
\bibitem [{\citenamefont {Koepernik}\ and\ \citenamefont
  {Eschrig}(1999)}]{Koepernik1999}%
  \BibitemOpen
  \bibfield  {author} {\bibinfo {author} {\bibfnamefont {K.}~\bibnamefont
  {Koepernik}}\ and\ \bibinfo {author} {\bibfnamefont {H.}~\bibnamefont
  {Eschrig}},\ }\href {https://link.aps.org/doi/10.1103/PhysRevB.59.1743}
  {\bibfield  {journal} {\bibinfo  {journal} {Physical Review B}\ }\textbf
  {\bibinfo {volume} {59}},\ \bibinfo {pages} {1743} (\bibinfo {year}
  {1999})}\BibitemShut {NoStop}%
\bibitem [{\citenamefont {Perdew}\ \emph {et~al.}(1996)\citenamefont {Perdew},
  \citenamefont {Burke},\ and\ \citenamefont {Ernzerhof}}]{Perdew1996}%
  \BibitemOpen
  \bibfield  {author} {\bibinfo {author} {\bibfnamefont {J.~P.}\ \bibnamefont
  {Perdew}}, \bibinfo {author} {\bibfnamefont {K.}~\bibnamefont {Burke}}, \
  and\ \bibinfo {author} {\bibfnamefont {M.}~\bibnamefont {Ernzerhof}},\ }\href
  {\doibase 10.1103/PhysRevLett.77.3865} {\bibfield  {journal} {\bibinfo
  {journal} {Physical Review Letters}\ }\textbf {\bibinfo {volume} {77}},\
  \bibinfo {pages} {3865} (\bibinfo {year} {1996})}\BibitemShut {NoStop}%
\bibitem [{\citenamefont {Mazurenko}\ \emph {et~al.}(2006)\citenamefont
  {Mazurenko}, \citenamefont {Mila},\ and\ \citenamefont
  {Anisimov}}]{Mazurenko2006}%
  \BibitemOpen
  \bibfield  {author} {\bibinfo {author} {\bibfnamefont {V.~V.}\ \bibnamefont
  {Mazurenko}}, \bibinfo {author} {\bibfnamefont {F.}~\bibnamefont {Mila}}, \
  and\ \bibinfo {author} {\bibfnamefont {V.~I.}\ \bibnamefont {Anisimov}},\
  }\href {\doibase 10.1103/PhysRevB.73.014418} {\bibfield  {journal} {\bibinfo
  {journal} {Physical Review B}\ }\textbf {\bibinfo {volume} {73}},\ \bibinfo
  {pages} {14418} (\bibinfo {year} {2006})}\BibitemShut {NoStop}%
\bibitem [{\citenamefont {Tsirlin}\ \emph
  {et~al.}(2011{\natexlab{a}})\citenamefont {Tsirlin}, \citenamefont {Janson},\
  and\ \citenamefont {Rosner}}]{Tsirlin2011}%
  \BibitemOpen
  \bibfield  {author} {\bibinfo {author} {\bibfnamefont {A.~A.}\ \bibnamefont
  {Tsirlin}}, \bibinfo {author} {\bibfnamefont {O.}~\bibnamefont {Janson}}, \
  and\ \bibinfo {author} {\bibfnamefont {H.}~\bibnamefont {Rosner}},\ }\href
  {\doibase 10.1103/PhysRevB.84.144429} {\bibfield  {journal} {\bibinfo
  {journal} {Physical Review B}\ }\textbf {\bibinfo {volume} {84}},\ \bibinfo
  {pages} {144429} (\bibinfo {year} {2011}{\natexlab{a}})}\BibitemShut
  {NoStop}%
\bibitem [{\citenamefont {Xiang}\ \emph {et~al.}(2011)\citenamefont {Xiang},
  \citenamefont {Kan}, \citenamefont {Wei}, \citenamefont {Whangbo},\ and\
  \citenamefont {Gong}}]{Xiang2011}%
  \BibitemOpen
  \bibfield  {author} {\bibinfo {author} {\bibfnamefont {H.~J.}\ \bibnamefont
  {Xiang}}, \bibinfo {author} {\bibfnamefont {E.~J.}\ \bibnamefont {Kan}},
  \bibinfo {author} {\bibfnamefont {S.-H.}\ \bibnamefont {Wei}}, \bibinfo
  {author} {\bibfnamefont {M.-H.}\ \bibnamefont {Whangbo}}, \ and\ \bibinfo
  {author} {\bibfnamefont {X.~G.}\ \bibnamefont {Gong}},\ }\href {\doibase
  10.1103/PhysRevB.84.224429} {\bibfield  {journal} {\bibinfo  {journal}
  {Physical Review B}\ }\textbf {\bibinfo {volume} {84}},\ \bibinfo {pages}
  {224429} (\bibinfo {year} {2011})}\BibitemShut {NoStop}%
\bibitem [{\citenamefont {Tsirlin}(2014)}]{Tsirlin2014}%
  \BibitemOpen
  \bibfield  {author} {\bibinfo {author} {\bibfnamefont {A.~A.}\ \bibnamefont
  {Tsirlin}},\ }\href {\doibase 10.1103/PhysRevB.89.014405} {\bibfield
  {journal} {\bibinfo  {journal} {Physical Review B}\ }\textbf {\bibinfo
  {volume} {89}},\ \bibinfo {pages} {14405} (\bibinfo {year}
  {2014})}\BibitemShut {NoStop}%
\bibitem [{\citenamefont {Iqbal}\ \emph {et~al.}(2017)\citenamefont {Iqbal},
  \citenamefont {Müller}, \citenamefont {Riedl}, \citenamefont {Reuther},
  \citenamefont {Rachel}, \citenamefont {Valentí}, \citenamefont {Gingras},
  \citenamefont {Thomale},\ and\ \citenamefont {Jeschke}}]{Iqbal2017}%
  \BibitemOpen
  \bibfield  {author} {\bibinfo {author} {\bibfnamefont {Y.}~\bibnamefont
  {Iqbal}}, \bibinfo {author} {\bibfnamefont {T.}~\bibnamefont {Müller}},
  \bibinfo {author} {\bibfnamefont {K.}~\bibnamefont {Riedl}}, \bibinfo
  {author} {\bibfnamefont {J.}~\bibnamefont {Reuther}}, \bibinfo {author}
  {\bibfnamefont {S.}~\bibnamefont {Rachel}}, \bibinfo {author} {\bibfnamefont
  {R.}~\bibnamefont {Valentí}}, \bibinfo {author} {\bibfnamefont {M.~J.~P.}\
  \bibnamefont {Gingras}}, \bibinfo {author} {\bibfnamefont {R.}~\bibnamefont
  {Thomale}}, \ and\ \bibinfo {author} {\bibfnamefont {H.~O.}\ \bibnamefont
  {Jeschke}},\ }\href {\doibase 10.1103/PhysRevMaterials.1.071201} {\bibfield
  {journal} {\bibinfo  {journal} {Physical Review Materials}\ }\textbf
  {\bibinfo {volume} {1}},\ \bibinfo {pages} {71201} (\bibinfo {year}
  {2017})}\BibitemShut {NoStop}%
\bibitem [{\citenamefont {Hembacher}\ \emph {et~al.}(2018)\citenamefont
  {Hembacher}, \citenamefont {Badrtdinov}, \citenamefont {Ding}, \citenamefont
  {Sobczak}, \citenamefont {Ritter}, \citenamefont {Mazurenko},\ and\
  \citenamefont {Tsirlin}}]{Hembacher2018}%
  \BibitemOpen
  \bibfield  {author} {\bibinfo {author} {\bibfnamefont {J.}~\bibnamefont
  {Hembacher}}, \bibinfo {author} {\bibfnamefont {D.~I.}\ \bibnamefont
  {Badrtdinov}}, \bibinfo {author} {\bibfnamefont {L.}~\bibnamefont {Ding}},
  \bibinfo {author} {\bibfnamefont {Z.}~\bibnamefont {Sobczak}}, \bibinfo
  {author} {\bibfnamefont {C.}~\bibnamefont {Ritter}}, \bibinfo {author}
  {\bibfnamefont {V.~V.}\ \bibnamefont {Mazurenko}}, \ and\ \bibinfo {author}
  {\bibfnamefont {A.~A.}\ \bibnamefont {Tsirlin}},\ }\href {\doibase
  10.1103/PhysRevB.98.094406} {\bibfield  {journal} {\bibinfo  {journal}
  {Physical Review B}\ }\textbf {\bibinfo {volume} {98}},\ \bibinfo {pages}
  {94406} (\bibinfo {year} {2018})}\BibitemShut {NoStop}%
\bibitem [{\citenamefont {Abdeldaim}\ \emph {et~al.}(2019)\citenamefont
  {Abdeldaim}, \citenamefont {Badrtdinov}, \citenamefont {Gibbs}, \citenamefont
  {Manuel}, \citenamefont {Walker}, \citenamefont {Le}, \citenamefont {Wu},
  \citenamefont {Wardecki}, \citenamefont {Eriksson}, \citenamefont {Kvashnin},
  \citenamefont {Tsirlin},\ and\ \citenamefont {Nilsen}}]{Abdeldaim2019}%
  \BibitemOpen
  \bibfield  {author} {\bibinfo {author} {\bibfnamefont {A.~H.}\ \bibnamefont
  {Abdeldaim}}, \bibinfo {author} {\bibfnamefont {D.~I.}\ \bibnamefont
  {Badrtdinov}}, \bibinfo {author} {\bibfnamefont {A.~S.}\ \bibnamefont
  {Gibbs}}, \bibinfo {author} {\bibfnamefont {P.}~\bibnamefont {Manuel}},
  \bibinfo {author} {\bibfnamefont {H.~C.}\ \bibnamefont {Walker}}, \bibinfo
  {author} {\bibfnamefont {M.~D.}\ \bibnamefont {Le}}, \bibinfo {author}
  {\bibfnamefont {C.~H.}\ \bibnamefont {Wu}}, \bibinfo {author} {\bibfnamefont
  {D.}~\bibnamefont {Wardecki}}, \bibinfo {author} {\bibfnamefont {S.-G.}\
  \bibnamefont {Eriksson}}, \bibinfo {author} {\bibfnamefont {Y.~O.}\
  \bibnamefont {Kvashnin}}, \bibinfo {author} {\bibfnamefont {A.~A.}\
  \bibnamefont {Tsirlin}}, \ and\ \bibinfo {author} {\bibfnamefont {G.~J.}\
  \bibnamefont {Nilsen}},\ }\href {\doibase 10.1103/PhysRevB.100.214427}
  {\bibfield  {journal} {\bibinfo  {journal} {Physical Review B}\ }\textbf
  {\bibinfo {volume} {100}},\ \bibinfo {pages} {214427} (\bibinfo {year}
  {2019})}\BibitemShut {NoStop}%
\bibitem [{\citenamefont {Urushihara}\ \emph {et~al.}(2020)\citenamefont
  {Urushihara}, \citenamefont {Kawaguchi}, \citenamefont {Fukuda},\ and\
  \citenamefont {Asaka}}]{Urushihara2020}%
  \BibitemOpen
  \bibfield  {author} {\bibinfo {author} {\bibfnamefont {D.}~\bibnamefont
  {Urushihara}}, \bibinfo {author} {\bibfnamefont {S.}~\bibnamefont
  {Kawaguchi}}, \bibinfo {author} {\bibfnamefont {K.}~\bibnamefont {Fukuda}}, \
  and\ \bibinfo {author} {\bibfnamefont {T.}~\bibnamefont {Asaka}},\ }\href
  {https://doi.org/10.1107/S2052252520005655} {\bibfield  {journal} {\bibinfo
  {journal} {IUCrJ}\ }\textbf {\bibinfo {volume} {7}},\ \bibinfo {pages} {656}
  (\bibinfo {year} {2020})}\BibitemShut {NoStop}%
\bibitem [{\citenamefont {de~Vries}\ \emph {et~al.}(2010)\citenamefont
  {de~Vries}, \citenamefont {Mclaughlin},\ and\ \citenamefont {Bos}}]{deVries}%
  \BibitemOpen
  \bibfield  {author} {\bibinfo {author} {\bibfnamefont {M.~A.}\ \bibnamefont
  {de~Vries}}, \bibinfo {author} {\bibfnamefont {A.~C.}\ \bibnamefont
  {Mclaughlin}}, \ and\ \bibinfo {author} {\bibfnamefont {J.-W.~G.}\
  \bibnamefont {Bos}},\ }\href {\doibase 10.1103/PhysRevLett.104.177202}
  {\bibfield  {journal} {\bibinfo  {journal} {Physical Review Letters}\
  }\textbf {\bibinfo {volume} {104}},\ \bibinfo {pages} {177202} (\bibinfo
  {year} {2010})}\BibitemShut {NoStop}%
\bibitem [{\citenamefont {Clark}\ \emph {et~al.}(2014)\citenamefont {Clark},
  \citenamefont {Nilsen}, \citenamefont {Kermarrec}, \citenamefont {Ehlers},
  \citenamefont {Knight}, \citenamefont {Harrison}, \citenamefont {Attfield},\
  and\ \citenamefont {Gaulin}}]{Clark2014}%
  \BibitemOpen
  \bibfield  {author} {\bibinfo {author} {\bibfnamefont {L.}~\bibnamefont
  {Clark}}, \bibinfo {author} {\bibfnamefont {G.}~\bibnamefont {Nilsen}},
  \bibinfo {author} {\bibfnamefont {E.}~\bibnamefont {Kermarrec}}, \bibinfo
  {author} {\bibfnamefont {G.}~\bibnamefont {Ehlers}}, \bibinfo {author}
  {\bibfnamefont {K.}~\bibnamefont {Knight}}, \bibinfo {author} {\bibfnamefont
  {A.}~\bibnamefont {Harrison}}, \bibinfo {author} {\bibfnamefont
  {J.}~\bibnamefont {Attfield}}, \ and\ \bibinfo {author} {\bibfnamefont
  {B.}~\bibnamefont {Gaulin}},\ }\href {\doibase
  10.1103/PhysRevLett.113.117201} {\bibfield  {journal} {\bibinfo  {journal}
  {Physical Review Letters}\ }\textbf {\bibinfo {volume} {113}},\ \bibinfo
  {pages} {117201} (\bibinfo {year} {2014})}\BibitemShut {NoStop}%
\bibitem [{\citenamefont {Johnston}\ \emph {et~al.}(2000)\citenamefont
  {Johnston}, \citenamefont {Kremer}, \citenamefont {Troyer}, \citenamefont
  {Wang}, \citenamefont {Klümper}, \citenamefont {Bud’ko}, \citenamefont
  {Panchula},\ and\ \citenamefont {Canfield}}]{Johnston2000}%
  \BibitemOpen
  \bibfield  {author} {\bibinfo {author} {\bibfnamefont {D.~C.}\ \bibnamefont
  {Johnston}}, \bibinfo {author} {\bibfnamefont {R.~K.}\ \bibnamefont
  {Kremer}}, \bibinfo {author} {\bibfnamefont {M.}~\bibnamefont {Troyer}},
  \bibinfo {author} {\bibfnamefont {X.}~\bibnamefont {Wang}}, \bibinfo {author}
  {\bibfnamefont {A.}~\bibnamefont {Klümper}}, \bibinfo {author}
  {\bibfnamefont {S.~L.}\ \bibnamefont {Bud’ko}}, \bibinfo {author}
  {\bibfnamefont {A.~F.}\ \bibnamefont {Panchula}}, \ and\ \bibinfo {author}
  {\bibfnamefont {P.~C.}\ \bibnamefont {Canfield}},\ }\href {\doibase
  10.1103/PhysRevB.61.9558} {\bibfield  {journal} {\bibinfo  {journal}
  {Physical Review B}\ }\textbf {\bibinfo {volume} {61}},\ \bibinfo {pages}
  {9558} (\bibinfo {year} {2000})}\BibitemShut {NoStop}%
\bibitem [{\citenamefont {Bernu}\ and\ \citenamefont
  {Misguich}(2001)}]{Bernu2001}%
  \BibitemOpen
  \bibfield  {author} {\bibinfo {author} {\bibfnamefont {B.}~\bibnamefont
  {Bernu}}\ and\ \bibinfo {author} {\bibfnamefont {G.}~\bibnamefont
  {Misguich}},\ }\href {\doibase 10.1103/PhysRevB.63.134409} {\bibfield
  {journal} {\bibinfo  {journal} {Physical Review B}\ }\textbf {\bibinfo
  {volume} {63}},\ \bibinfo {pages} {134409} (\bibinfo {year}
  {2001})}\BibitemShut {NoStop}%
\bibitem [{\citenamefont {Schulz}(1996)}]{Schulz1996}%
  \BibitemOpen
  \bibfield  {author} {\bibinfo {author} {\bibfnamefont {H.~J.}\ \bibnamefont
  {Schulz}},\ }\href {\doibase 10.1103/PhysRevLett.77.2790} {\bibfield
  {journal} {\bibinfo  {journal} {Physical Review Letters}\ }\textbf {\bibinfo
  {volume} {77}},\ \bibinfo {pages} {2790} (\bibinfo {year}
  {1996})}\BibitemShut {NoStop}%
\bibitem [{\citenamefont {Nilsen}\ \emph {et~al.}(2015)\citenamefont {Nilsen},
  \citenamefont {Raja}, \citenamefont {Tsirlin}, \citenamefont {Mutka},
  \citenamefont {Kasinathan}, \citenamefont {Ritter},\ and\ \citenamefont
  {Rønnow}}]{Nilsen2015}%
  \BibitemOpen
  \bibfield  {author} {\bibinfo {author} {\bibfnamefont {G.~J.}\ \bibnamefont
  {Nilsen}}, \bibinfo {author} {\bibfnamefont {A.}~\bibnamefont {Raja}},
  \bibinfo {author} {\bibfnamefont {A.~A.}\ \bibnamefont {Tsirlin}}, \bibinfo
  {author} {\bibfnamefont {H.}~\bibnamefont {Mutka}}, \bibinfo {author}
  {\bibfnamefont {D.}~\bibnamefont {Kasinathan}}, \bibinfo {author}
  {\bibfnamefont {C.}~\bibnamefont {Ritter}}, \ and\ \bibinfo {author}
  {\bibfnamefont {H.~M.}\ \bibnamefont {Rønnow}},\ }\href
  {http://dx.doi.org/10.1088/1367-2630/17/11/113035} {\bibfield  {journal}
  {\bibinfo  {journal} {New Journal of Physics}\ }\textbf {\bibinfo {volume}
  {17}},\ \bibinfo {pages} {113035} (\bibinfo {year} {2015})}\BibitemShut
  {NoStop}%
\bibitem [{\citenamefont {Belik}\ \emph {et~al.}(2005)\citenamefont {Belik},
  \citenamefont {Uji}, \citenamefont {Terashima},\ and\ \citenamefont
  {Takayama-Muromachi}}]{Belik2005}%
  \BibitemOpen
  \bibfield  {author} {\bibinfo {author} {\bibfnamefont {A.~A.}\ \bibnamefont
  {Belik}}, \bibinfo {author} {\bibfnamefont {S.}~\bibnamefont {Uji}}, \bibinfo
  {author} {\bibfnamefont {T.}~\bibnamefont {Terashima}}, \ and\ \bibinfo
  {author} {\bibfnamefont {E.}~\bibnamefont {Takayama-Muromachi}},\ }\href
  {https://www.sciencedirect.com/science/article/pii/S0022459605003841}
  {\bibfield  {journal} {\bibinfo  {journal} {Journal of Solid State
  Chemistry}\ }\textbf {\bibinfo {volume} {178}},\ \bibinfo {pages} {3461}
  (\bibinfo {year} {2005})}\BibitemShut {NoStop}%
\bibitem [{\citenamefont {Mukharjee}\ \emph {et~al.}(2021)\citenamefont
  {Mukharjee}, \citenamefont {Somesh}, \citenamefont {Ranjith}, \citenamefont
  {Baenitz}, \citenamefont {Skourski}, \citenamefont {Adroja}, \citenamefont
  {Khalyavin}, \citenamefont {Tsirlin},\ and\ \citenamefont
  {Nath}}]{Mukharjee2021}%
  \BibitemOpen
  \bibfield  {author} {\bibinfo {author} {\bibfnamefont {P.~K.}\ \bibnamefont
  {Mukharjee}}, \bibinfo {author} {\bibfnamefont {K.}~\bibnamefont {Somesh}},
  \bibinfo {author} {\bibfnamefont {K.~M.}\ \bibnamefont {Ranjith}}, \bibinfo
  {author} {\bibfnamefont {M.}~\bibnamefont {Baenitz}}, \bibinfo {author}
  {\bibfnamefont {Y.}~\bibnamefont {Skourski}}, \bibinfo {author}
  {\bibfnamefont {D.~T.}\ \bibnamefont {Adroja}}, \bibinfo {author}
  {\bibfnamefont {D.}~\bibnamefont {Khalyavin}}, \bibinfo {author}
  {\bibfnamefont {A.~A.}\ \bibnamefont {Tsirlin}}, \ and\ \bibinfo {author}
  {\bibfnamefont {R.}~\bibnamefont {Nath}},\ }\href
  {https://link.aps.org/doi/10.1103/PhysRevB.104.224409} {\bibfield  {journal}
  {\bibinfo  {journal} {Physical Review B}\ }\textbf {\bibinfo {volume}
  {104}},\ \bibinfo {pages} {224409} (\bibinfo {year} {2021})}\BibitemShut
  {NoStop}%
\bibitem [{\citenamefont {Tsirlin}\ \emph
  {et~al.}(2011{\natexlab{b}})\citenamefont {Tsirlin}, \citenamefont {Nath},
  \citenamefont {Sichelschmidt}, \citenamefont {Skourski}, \citenamefont
  {Geibel},\ and\ \citenamefont {Rosner}}]{Tsirlin2011_2}%
  \BibitemOpen
  \bibfield  {author} {\bibinfo {author} {\bibfnamefont {A.~A.}\ \bibnamefont
  {Tsirlin}}, \bibinfo {author} {\bibfnamefont {R.}~\bibnamefont {Nath}},
  \bibinfo {author} {\bibfnamefont {J.}~\bibnamefont {Sichelschmidt}}, \bibinfo
  {author} {\bibfnamefont {Y.}~\bibnamefont {Skourski}}, \bibinfo {author}
  {\bibfnamefont {C.}~\bibnamefont {Geibel}}, \ and\ \bibinfo {author}
  {\bibfnamefont {H.}~\bibnamefont {Rosner}},\ }\href {\doibase
  10.1103/PhysRevB.83.144412} {\bibfield  {journal} {\bibinfo  {journal}
  {Physical Review B}\ }\textbf {\bibinfo {volume} {83}},\ \bibinfo {pages}
  {144412} (\bibinfo {year} {2011}{\natexlab{b}})}\BibitemShut {NoStop}%
\bibitem [{\citenamefont {Garrett}\ \emph {et~al.}(1997)\citenamefont
  {Garrett}, \citenamefont {Nagler}, \citenamefont {Tennant}, \citenamefont
  {Sales},\ and\ \citenamefont {Barnes}}]{Garrett1997}%
  \BibitemOpen
  \bibfield  {author} {\bibinfo {author} {\bibfnamefont {A.~W.}\ \bibnamefont
  {Garrett}}, \bibinfo {author} {\bibfnamefont {S.~E.}\ \bibnamefont {Nagler}},
  \bibinfo {author} {\bibfnamefont {D.~A.}\ \bibnamefont {Tennant}}, \bibinfo
  {author} {\bibfnamefont {B.~C.}\ \bibnamefont {Sales}}, \ and\ \bibinfo
  {author} {\bibfnamefont {T.}~\bibnamefont {Barnes}},\ }\href
  {https://link.aps.org/doi/10.1103/PhysRevLett.79.745} {\bibfield  {journal}
  {\bibinfo  {journal} {Physical Review Letters}\ }\textbf {\bibinfo {volume}
  {79}},\ \bibinfo {pages} {745} (\bibinfo {year} {1997})}\BibitemShut
  {NoStop}%
\bibitem [{\citenamefont {Tennant}\ \emph {et~al.}(1997)\citenamefont
  {Tennant}, \citenamefont {Nagler}, \citenamefont {Garrett}, \citenamefont
  {Barnes},\ and\ \citenamefont {Torardi}}]{Tennant1997}%
  \BibitemOpen
  \bibfield  {author} {\bibinfo {author} {\bibfnamefont {D.~A.}\ \bibnamefont
  {Tennant}}, \bibinfo {author} {\bibfnamefont {S.~E.}\ \bibnamefont {Nagler}},
  \bibinfo {author} {\bibfnamefont {A.~W.}\ \bibnamefont {Garrett}}, \bibinfo
  {author} {\bibfnamefont {T.}~\bibnamefont {Barnes}}, \ and\ \bibinfo {author}
  {\bibfnamefont {C.~C.}\ \bibnamefont {Torardi}},\ }\href
  {https://link.aps.org/doi/10.1103/PhysRevLett.78.4998} {\bibfield  {journal}
  {\bibinfo  {journal} {Physical Review Letters}\ }\textbf {\bibinfo {volume}
  {78}},\ \bibinfo {pages} {4998} (\bibinfo {year} {1997})}\BibitemShut
  {NoStop}%
\bibitem [{\citenamefont {Roca}\ \emph {et~al.}(1998)\citenamefont {Roca},
  \citenamefont {Amorós}, \citenamefont {Cano}, \citenamefont {Marcos},
  \citenamefont {Alamo}, \citenamefont {Beltrán-Porter},\ and\ \citenamefont
  {Beltrán-Porter}}]{Roca1998}%
  \BibitemOpen
  \bibfield  {author} {\bibinfo {author} {\bibfnamefont {M.}~\bibnamefont
  {Roca}}, \bibinfo {author} {\bibfnamefont {P.}~\bibnamefont {Amorós}},
  \bibinfo {author} {\bibfnamefont {J.}~\bibnamefont {Cano}}, \bibinfo {author}
  {\bibfnamefont {M.~D.}\ \bibnamefont {Marcos}}, \bibinfo {author}
  {\bibfnamefont {J.}~\bibnamefont {Alamo}}, \bibinfo {author} {\bibfnamefont
  {A.}~\bibnamefont {Beltrán-Porter}}, \ and\ \bibinfo {author} {\bibfnamefont
  {D.}~\bibnamefont {Beltrán-Porter}},\ }\href {\doibase 10.1021/ic971210o}
  {\bibfield  {journal} {\bibinfo  {journal} {Inorganic Chemistry}\ }\textbf
  {\bibinfo {volume} {37}},\ \bibinfo {pages} {3167} (\bibinfo {year}
  {1998})}\BibitemShut {NoStop}%
\bibitem [{\citenamefont {H\"alg}\ \emph {et~al.}(2014)\citenamefont {H\"alg},
  \citenamefont {Lorenz}, \citenamefont {Povarov}, \citenamefont {M\aa{}nsson},
  \citenamefont {Skourski},\ and\ \citenamefont {Zheludev}}]{dm_exp}%
  \BibitemOpen
  \bibfield  {author} {\bibinfo {author} {\bibfnamefont {M.}~\bibnamefont
  {H\"alg}}, \bibinfo {author} {\bibfnamefont {W.~E.~A.}\ \bibnamefont
  {Lorenz}}, \bibinfo {author} {\bibfnamefont {K.~Y.}\ \bibnamefont {Povarov}},
  \bibinfo {author} {\bibfnamefont {M.}~\bibnamefont {M\aa{}nsson}}, \bibinfo
  {author} {\bibfnamefont {Y.}~\bibnamefont {Skourski}}, \ and\ \bibinfo
  {author} {\bibfnamefont {A.}~\bibnamefont {Zheludev}},\ }\href {\doibase
  10.1103/PhysRevB.90.174413} {\bibfield  {journal} {\bibinfo  {journal} {Phys.
  Rev. B}\ }\textbf {\bibinfo {volume} {90}},\ \bibinfo {pages} {174413}
  (\bibinfo {year} {2014})}\BibitemShut {NoStop}%
\bibitem [{\citenamefont {Wang}\ \emph {et~al.}(2022)\citenamefont {Wang},
  \citenamefont {Keselman},\ and\ \citenamefont {Starykh}}]{dm_THEORY}%
  \BibitemOpen
  \bibfield  {author} {\bibinfo {author} {\bibfnamefont {R.-B.}\ \bibnamefont
  {Wang}}, \bibinfo {author} {\bibfnamefont {A.}~\bibnamefont {Keselman}}, \
  and\ \bibinfo {author} {\bibfnamefont {O.~A.}\ \bibnamefont {Starykh}},\
  }\href {\doibase 10.1103/PhysRevB.105.184429} {\bibfield  {journal} {\bibinfo
   {journal} {Phys. Rev. B}\ }\textbf {\bibinfo {volume} {105}},\ \bibinfo
  {pages} {184429} (\bibinfo {year} {2022})}\BibitemShut {NoStop}%
\bibitem [{\citenamefont {Janson}\ \emph {et~al.}(2009)\citenamefont {Janson},
  \citenamefont {Schnelle}, \citenamefont {Schmidt}, \citenamefont {Prots},
  \citenamefont {Drechsler}, \citenamefont {Filatov},\ and\ \citenamefont
  {Rosner}}]{Janson2009}%
  \BibitemOpen
  \bibfield  {author} {\bibinfo {author} {\bibfnamefont {O.}~\bibnamefont
  {Janson}}, \bibinfo {author} {\bibfnamefont {W.}~\bibnamefont {Schnelle}},
  \bibinfo {author} {\bibfnamefont {M.}~\bibnamefont {Schmidt}}, \bibinfo
  {author} {\bibfnamefont {Y.}~\bibnamefont {Prots}}, \bibinfo {author}
  {\bibfnamefont {S.-L.}\ \bibnamefont {Drechsler}}, \bibinfo {author}
  {\bibfnamefont {S.~K.}\ \bibnamefont {Filatov}}, \ and\ \bibinfo {author}
  {\bibfnamefont {H.}~\bibnamefont {Rosner}},\ }\href {\doibase
  10.1088/1367-2630/11/11/113034} {\bibfield  {journal} {\bibinfo  {journal}
  {New Journal of Physics}\ }\textbf {\bibinfo {volume} {11}},\ \bibinfo
  {pages} {113034} (\bibinfo {year} {2009})}\BibitemShut {NoStop}%
\bibitem [{\citenamefont {Kojima}\ \emph {et~al.}(1997)\citenamefont {Kojima},
  \citenamefont {Fudamoto}, \citenamefont {Larkin}, \citenamefont {Luke},
  \citenamefont {Merrin}, \citenamefont {Nachumi}, \citenamefont {Uemura},
  \citenamefont {Motoyama}, \citenamefont {Eisaki}, \citenamefont {Uchida},
  \citenamefont {Yamada}, \citenamefont {Endoh}, \citenamefont {Hosoya},
  \citenamefont {Sternlieb},\ and\ \citenamefont {Shirane}}]{Kojima1997}%
  \BibitemOpen
  \bibfield  {author} {\bibinfo {author} {\bibfnamefont {K.~M.}\ \bibnamefont
  {Kojima}}, \bibinfo {author} {\bibfnamefont {Y.}~\bibnamefont {Fudamoto}},
  \bibinfo {author} {\bibfnamefont {M.}~\bibnamefont {Larkin}}, \bibinfo
  {author} {\bibfnamefont {G.~M.}\ \bibnamefont {Luke}}, \bibinfo {author}
  {\bibfnamefont {J.}~\bibnamefont {Merrin}}, \bibinfo {author} {\bibfnamefont
  {B.}~\bibnamefont {Nachumi}}, \bibinfo {author} {\bibfnamefont {Y.~J.}\
  \bibnamefont {Uemura}}, \bibinfo {author} {\bibfnamefont {N.}~\bibnamefont
  {Motoyama}}, \bibinfo {author} {\bibfnamefont {H.}~\bibnamefont {Eisaki}},
  \bibinfo {author} {\bibfnamefont {S.}~\bibnamefont {Uchida}}, \bibinfo
  {author} {\bibfnamefont {K.}~\bibnamefont {Yamada}}, \bibinfo {author}
  {\bibfnamefont {Y.}~\bibnamefont {Endoh}}, \bibinfo {author} {\bibfnamefont
  {S.}~\bibnamefont {Hosoya}}, \bibinfo {author} {\bibfnamefont {B.~J.}\
  \bibnamefont {Sternlieb}}, \ and\ \bibinfo {author} {\bibfnamefont
  {G.}~\bibnamefont {Shirane}},\ }\href {\doibase 10.1103/PhysRevLett.78.1787}
  {\bibfield  {journal} {\bibinfo  {journal} {Physical Review Letters}\
  }\textbf {\bibinfo {volume} {78}},\ \bibinfo {pages} {1787} (\bibinfo {year}
  {1997})}\BibitemShut {NoStop}%
\bibitem [{\citenamefont {Rosner}\ \emph {et~al.}(1997)\citenamefont {Rosner},
  \citenamefont {Eschrig}, \citenamefont {Hayn}, \citenamefont {Drechsler},\
  and\ \citenamefont {Málek}}]{Rosner1997}%
  \BibitemOpen
  \bibfield  {author} {\bibinfo {author} {\bibfnamefont {H.}~\bibnamefont
  {Rosner}}, \bibinfo {author} {\bibfnamefont {H.}~\bibnamefont {Eschrig}},
  \bibinfo {author} {\bibfnamefont {R.}~\bibnamefont {Hayn}}, \bibinfo {author}
  {\bibfnamefont {S.-L.}\ \bibnamefont {Drechsler}}, \ and\ \bibinfo {author}
  {\bibfnamefont {J.}~\bibnamefont {Málek}},\ }\href {\doibase
  10.1103/PhysRevB.56.3402} {\bibfield  {journal} {\bibinfo  {journal}
  {Physical Review B}\ }\textbf {\bibinfo {volume} {56}},\ \bibinfo {pages}
  {3402} (\bibinfo {year} {1997})}\BibitemShut {NoStop}%
\bibitem [{\citenamefont {Paidi}\ \emph {et~al.}(2015)\citenamefont {Paidi},
  \citenamefont {Devi},\ and\ \citenamefont {Vidyasagar}}]{Paidi2015}%
  \BibitemOpen
  \bibfield  {author} {\bibinfo {author} {\bibfnamefont {A.~K.}\ \bibnamefont
  {Paidi}}, \bibinfo {author} {\bibfnamefont {R.~N.}\ \bibnamefont {Devi}}, \
  and\ \bibinfo {author} {\bibfnamefont {K.}~\bibnamefont {Vidyasagar}},\
  }\href {http://dx.doi.org/10.1039/C5DT02618K} {\bibfield  {journal} {\bibinfo
   {journal} {Dalton Transactions}\ }\textbf {\bibinfo {volume} {44}},\
  \bibinfo {pages} {17399} (\bibinfo {year} {2015})}\BibitemShut {NoStop}%
\bibitem [{\citenamefont {Leclaire}\ \emph {et~al.}(2001)\citenamefont
  {Leclaire}, \citenamefont {Chardon},\ and\ \citenamefont
  {Raveau}}]{Leclaire2001}%
  \BibitemOpen
  \bibfield  {author} {\bibinfo {author} {\bibfnamefont {A.}~\bibnamefont
  {Leclaire}}, \bibinfo {author} {\bibfnamefont {J.}~\bibnamefont {Chardon}}, \
  and\ \bibinfo {author} {\bibfnamefont {B.}~\bibnamefont {Raveau}},\ }\href
  {http://dx.doi.org/10.1039/B007592M} {\bibfield  {journal} {\bibinfo
  {journal} {Journal of Materials Chemistry}\ }\textbf {\bibinfo {volume}
  {11}},\ \bibinfo {pages} {846} (\bibinfo {year} {2001})}\BibitemShut
  {NoStop}%
\end{thebibliography}%

\end{document}